	\newcommand\Hom{{\rm Hom}}
	\newcommand{\ban}{\begin{eqnarray*}}
        \newcommand{\ean}{\end{eqnarray*}}
	\def\section#1{\vskip3em{\centerline {\bf#1}}\vskip3em}

        \newcommand{\ee}{\end{equation}}
        \newcommand{\ba}{\begin{eqnarray}}
        \newcommand{\eps}{\epsilon}
        
	\newcommand{\om}{\omega}
	\newcommand{\Om}{\Omega}
	\newcommand{\Chi}{{\cal X}}

        \newcommand{\tensor}{\otimes}
        \newcommand{\maps}{\colon}
        \newcommand{\ea}{\end{eqnarray}}
        \newcommand{\A}{{\cal A}}
        \newcommand{\G}{{\cal G}}
	
	\newcommand{\M}{{\cal M}}
	\renewcommand{\P}{{\cal P}}
	\newcommand{\D}{{\cal D}}
	\newcommand{\F}{{\cal F}}
	\renewcommand{\H}{{\bf H}}
	\newcommand{\I}{{\bf I}}
	\newcommand{\J}{{\bf J}}
	\newcommand{\K}{{\bf K}}
        \newcommand{\T}{{\cal T}}

        \newcommand{\g}{{\bf  g}}	%should be gothic
	\newcommand{\so}{{\bf so}}      %should be gothic
	\newcommand{\R}{{\bf R}}
	\newcommand{\C}{{\bf C}}
	\newcommand{\Z}{{\bf Z}}

	\newcommand{\hf}{{1\over 2}}
	\newcommand{\iso}{\cong}
	
	\newcommand{\Ad}{{\rm Ad}}
	\newcommand{\tr}{{\rm tr}}

	\newcommand{\Diff}{{\rm Diff}}
	\newcommand{\Maps}{{\rm Maps}}
	\newcommand{\End}{{\rm End}}
	\newcommand{\Fun}{{\rm Fun}}

	\documentstyle[12pt]{article}
\begin{document}
	\begin{center}
	{\bf Strings, Loops, Knots and Gauge Fields\\ }

        \vspace{0.5cm}
	{\em John C. Baez\\}
	\vspace{0.3cm}
	{\small Department of Mathematics \\
	University of California  \\
	Riverside CA 92521\\ }
	\vspace{0.3cm}
	{\small September 10, 1993\\ }
        \vspace{0.3 cm}
        {\small to appear in {\sl Knots and Quantum Gravity,} \\
        ed.\ J.\ Baez, Oxford U.\ Press\\ }
	\vspace{0.5cm}
	\end{center}

	\begin{abstract}
The loop representation of quantum gravity has many formal
resemblances to a background-free string theory.
In fact, its origins lie in attempts to treat the string theory
of hadrons as an approximation to QCD, in which the strings represent
flux tubes of the gauge field.  A heuristic path-integral approach
indicates a duality between background-free string theories and generally
covariant gauge theories, with the loop transform relating the two.
We review progress towards making this duality rigorous in three
examples: 2d Yang-Mills theory (which, while not generally covariant,
has symmetry under all area-preserving transformations), 3d quantum
gravity, and 4d quantum gravity.   $SU(N)$ Yang-Mills theory in 2
dimensions has been given a string-theoretic interpretation in the
large-$N$ limit by Gross, Taylor, Minahan and Polychronakos, but here
we provide an exact string-theoretic interpretation of the theory on
$\R\times S^1$ for finite $N$.  The string-theoretic
interpretation of quantum gravity in 3 dimensions gives rise to
conjectures about integrals on the moduli space of flat connections,
while in 4 dimensions there may be connections to the theory of 2-tangles.
\end{abstract}

\section{Introduction}

The notion of a deep relationship between string theories
and gauge theories is far from new.  String theory first arose
as a model of hadron interactions.
Unfortunately this theory had a number of undesirable
features; in particular, it predicted massless spin-2 particles.
It was soon supplanted by quantum chromodynamics (QCD), which models the
strong force by an $SU(3)$ Yang-Mills field.   However, string
models continued to be popular as an approximation of the confining phase
of QCD.  Two quarks in a meson, for
example, can be thought of as connected by a string-like flux tube in
which the gauge field is concentrated, while
an excitation of the gauge field alone can be thought of as a looped
flux tube.   This is essentially a modern reincarnation of
Faraday's notion of ``field lines,'' but it can be formalized using the
notion of Wilson loops.
If $A$ denotes a classical gauge field, or connection, a Wilson
loop is simply the trace of the holonomy of $A$ around a loop $\gamma$ in
space, typically written in terms of a path-ordered exponential
\[           \tr\, {\rm P}\, e^{ \oint_\gamma A }. \]
If instead $A$ denotes a quantized gauge field, the Wilson loop may be
reinterpreted as an operator on the Hilbert space of states, and
applying this operator to the vacuum state one obtains a state
in which the Yang-Mills analog of the electric field flows around
the loop $\gamma$.

In the late 1970's, Makeenko and Migdal, Nambu,
Polyakov, and others \cite{Nambu,Polyakov} attempted to derive
equations of string dynamics as an approximation to
the Yang-Mills equation, using Wilson loops.   More
recently, D.\ Gross and others
\cite{Gross,GT,Minahan,MP,NRS} have been able to {\it exactly} reformulate
Yang-Mills theory in 2-dimensional spacetime as a string theory
by writing an asymptotic series for the
vacuum expectation values of Wilson loops
as a sum over maps from surfaces (the string
worldsheet) to spacetime.   This development raises the hope that other
gauge theories might also be isomorphic to string theories.  For
example, recent work
by Witten \cite{Witten3} and Periwal \cite{Periwal} suggests that
Chern-Simons theory in 3 dimensions is also equivalent to a string
theory.

String theory eventually became popular as a theory of everything
because the massless spin-2 particles it predicted could be interpreted
as the gravitons one obtains by quantizing the spacetime
metric perturbatively about a fixed ``background'' metric.  Since
string theory appears to avoid the renormalization problems in
perturbative quantum gravity, it is a strong candidate for a theory
unifying gravity with the other forces.
However, while classical general relativity is an elegant
geometrical theory relying on no background structure for its formulation,
it has proved difficult to describe string theory along these lines.
Typically one begins with a fixed background structure and writes down
a string field theory in terms of this;
only afterwards can one investigate its background
independence \cite{Zwiebach}.
The clarity of a manifestly background-free approach
to string theory would be highly desirable.

On the other hand, attempts to formulate Yang-Mills theory in terms of
Wilson loops eventually led to a full-fledged ``loop representation'' of
gauge theories, thanks to the work of Gambini, Trias \cite{GamTri}, and
others.   After Ashtekar \cite{A} formulated quantum gravity as a sort of gauge
theory using the ``new variables,''  Rovelli and Smolin \cite{RS}
were able to use
the loop representation to study quantum gravity nonperturbatively in a
manifestly background-free formalism.   While superficially
quite different from modern string theory, this approach to quantum gravity
has many points of similarity, thanks to its common origin.  In particular,
it uses the device of Wilson loops to construct a space of states
consisting of ``multiloop invariants,'' which assign an amplitude to any
collection of loops in space.  The resemblance of these states to
wavefunctions of a string field theory is striking.  It is natural,
therefore, to ask whether the loop representation of quantum gravity might
be a string theory in disguise - or vice versa.

The present paper does not attempt a definitive answer to this question.
Rather, we begin by describing a general framework relating gauge theories and
string theories, and then consider a variety of examples.  Our treatment of
examples is also meant to serve as a review of
Yang-Mills theory in 2 dimensions and quantum gravity in 3 and 4 dimensions.

In Section 2 we describe how the loop
representation of a generally covariant gauge theories is related to a
background-free closed string field theory. We take a very naive approach
to strings, thinking of them simply as maps from a surface into spacetime,
and disregarding any conformal structure or fields propagating on the
surface. We base our treatment on the path integral formalism, and in
order to simplify the presentation we make a number of over-optimistic
assumptions concerning measures on infinite-dimensional spaces such as the
space $\A/\G$ of connections modulo gauge transformations.

In Section 3 we consider Yang-Mills theory in 2 dimensions as an example.
In fact, this theory is not generally covariant, but it has an
infinite-dimensional subgroup of the diffeomorphism group as symmetries,
the group of all area-preserving transformations.   Rather than the
path-integral approach we use canonical quantization, which is easier to
make rigorous.   Gross, Taylor, Minahan, and Polychronakos
\cite{Gross,GT,Minahan,MP}  have already given 2-dimensional $SU(N)$
Yang-Mills theory a string-theoretic interpretation in the large $N$ limit.
Our treatment is mostly a review of their work, but we find it to be little
extra effort, and rather enlightening, to give the theory a precise
string-theoretic interpretation for finite $N$.

In Section 4 we consider quantum gravity in 3 dimensions.  We review the
loop representation of this theory and raise some questions about integrals
over the moduli space of flat connections on a Riemann surface whose
resolution would be desirable for developing a string-theoretic picture of
the theory.  We also briefly discuss Chern-Simons theory in 3 dimensions.

These examples have finite-dimensional reduced configuration spaces, so
there are no analytical difficulties with measures on infinite-dimensional
spaces, at least in canonical quantization. In Section 5, however, we
consider quantum gravity in 4 dimensions. Here the classical configuration
space is infinite-dimensional and issues of analysis become more important.
We review recent work by Ashtekar, Isham, Lewandowski and the author
\cite{AI,AL,Baez} on diffeomorphism-invariant generalized measures on
$\A/\G$ and their relation to multiloop invariants and knot theory.  We
also note how a string-theoretic interpretation of the theory leads
naturally to the study of 2-tangles.

{\it Acknowledgements.} I would like to thank Abhay Ashtekar, Scott
Axelrod, Scott Carter, Paolo Cotta-Ramusino, Louis Crane, Jacob Hirbawi,
Jerzy Lewandowski, Renate Loll,  Maurizio Martellini, Jorge Pullin,
Holger Nielsen, and Lee Smolin for useful discussions.   Wati Taylor
deserves special thanks for explaining his work on Yang-Mills theory to me.
Also, I would like to collectively thank the Center for Gravitational
Physics and Geometry for inviting me to speak on this subject.

\section{String Field/Gauge Field Duality}

In this section we sketch a relationship between  string field theories and
gauge theories.   We begin with a nonperturbative Lagrangian description of
background-free closed string field theories. From this we derive a
Hamiltonian description, which turns out to be mathematically isomorphic to
the loop representation of a generally covariant gauge theory.   We
emphasize that while our discussion here is rigorous, it is schematic, in
the sense that some of our assumptions are not likely to hold precisely as
stated in the most interesting examples.   In particular, by ``measure'' in
this section we will always mean a positive regular Borel measure,  but in
fact one should work with a more general version of this concept.   We
discuss these analytical issues more carefully in Section 5.

Consider a theory of strings
propagating on a spacetime $M$ that is diffeomorphic to
$\R\times X$, with $X$ a manifold we call ``space.'' We do not
assume a {\it canonical} identification of $M$ with
$\R\times X$, or any other background structure (metric, etc.) on
spacetime.  We take the classical
configuration space of the string theory to be the space $\M$
of multiloops in $X$:
\[ \M = \bigcup_{n \ge 0} \M_n \]
with
\[ \M_n = \Maps(nS^1, X) .\]
Here $nS^1$ denotes the disjoint union of $n$ copies
of $S^1$, and we write ``Maps'' to denote the set of maps satisfying
some regularity conditions (continuity, smoothness, etc.) to be
specified.  Let $\D\gamma$
denote a measure on $\M$ and let
$\Fun(\M)$ denote some space of square-integrable functions on $\M$.
We assume that $\Fun(\M)$ and the measure
$\D\gamma$ are invariant both under diffeomorphisms of space and
reparametrizations of the strings.  That is, both the
identity component of the diffeomorphism group of $X$ and
the orientation-preserving diffeomorphisms of $nS^1$
act on $\M$, and we wish $\Fun(\M)$ and $\D\gamma$ to be preserved by
these actions.

Introduce on $\Fun(M)$ the ``kinematical inner product,'' which is
just the $L^2$ inner product
\[ \langle \psi,\phi\rangle_{kin} = \int_{\M}
\overline\psi(\gamma) \phi(\gamma) \, \D\gamma  .\]
We assume for convenience that this really is an inner product, i.e.\
it is nondegenerate.
Define the ``kinematical state space''
$\H_{kin}$ to be the Hilbert space completion of
$\Fun(\M)$ in the norm associated to this inner product.

Following ideas from canonical quantum gravity, we do not expect $\H_{kin}$
to be the true space of physical states.   In the space of physical states,
any two states differing by a diffeomorphism of spacetime are identified.
The physical state space thus depends on the dynamics of the theory.
Taking a Lagrangian approach, dynamics may be described using in terms of
path integrals as follows.   Fix a time $T > 0$.  Let $\P$ denote the set
of ``histories,'' that is, maps $f \maps \Sigma \to [0,T] \times X$, where
$\Sigma$ is a compact oriented 2-manifold with boundary, such that
\[       f(\Sigma) \cap \partial([0,T] \times X) = f(\partial\Sigma) .\]
Given
$\gamma,\gamma' \in \M$, we say that $f \in \P$ is a history from $\gamma$
to $\gamma'$ if $f \maps \Sigma \to [0,T] \times X$ and the boundary of
$\Sigma$ is a disjoint union of circles $nS^1 \cup mS^1$, with
 \[          f|nS^1 = \gamma ,\qquad f|mS^1 = \gamma' .\]
We fix a measure, or ``path
integral,'' on $\P(\gamma,\gamma')$.  Following tradition, we write this as
$ e^{iS(f)} \D f$, with $S(f)$ denoting the action of $f$, but $e^{iS(f)}$
and $\D f$ only appear in the combination $ e^{iS(f)}\D f$.  Since we are
interested in generally covariant theories, this path integral is assumed
to have some invariance properties, which we note below as they are needed.

Using the standard recipe in topological quantum field theory, we define
the ``physical inner product'' on $\H_{kin}$ by
\[ \langle \psi,\phi \rangle_{phys} = \int_{\M} \int_{\M}
 \int_{\P(\gamma,\gamma')}
\overline\psi(\gamma) \phi(\gamma') \, e^{iS(f)} \D f \D\gamma \D\gamma'\]
assuming optimistically that this integral is well-defined.
We do not actually assume this is an inner product in the standard
sense, for while we assume
$\langle \psi,\psi \rangle \ge 0$ for all $\psi \in \H_{kin}$, we do not
assume positive definiteness.
The general covariance of the theory should imply that this inner
product is independent of the choice of time $T > 0$, so we assume this
as well.

Define the space of norm-zero states $\I \subseteq \H_{kin}$ by
\ban        \I &=& \{ \psi | \; \langle\psi,\psi\rangle_{phys} = 0\} \\
&=&   \{ \psi | \;  \langle\psi,\phi \rangle_{phys}
= 0 \;{\rm for \; all }\; \phi \in \H_{kin} \}  \ean
and define the ``physical state space'' $\H_{phys}$ to be the Hilbert
space completion of $\H_{kin}/\I$ in the norm associated to the
physical inner product.
In general $\I$ is nonempty, because if $g \in \Diff_0(X)$ is a
diffeomorphism in the connected component of the identity,
we can find a path of diffeomorphisms
$g_t \in \Diff_0(M)$ with $g_0 = g$ and $g_T$ equal to the identity,
and defining
$\widetilde g \in \Diff([0,T] \times X)$ by
\[        \widetilde g(t,x) = (t,g_t(x)) ,\]
we have
\ban   \langle \psi,\phi\rangle_{phys} &=&
 \int_{\M} \int_{\M}
 \int_{\P(\gamma,\gamma')}
\overline\psi(\gamma) \phi(\gamma')e^{iS(f)}\, \D f  \D\gamma\D\gamma' \\
&=&   \int_{\M}  \int_{\M} \int_{\P(\gamma,\gamma')}
\overline{g\psi}(g\gamma) \phi(\gamma')\,  e^{iS(f)}  \D f\D\gamma\D\gamma' \\
&=&   \int_{\M}  \int_{\M}  \int_{\P(\gamma,\gamma')}
\overline{g\psi}(g\gamma) \phi(\gamma')\,  e^{iS(\tilde gf)}\, \D
(\widetilde gf)\D(g\gamma) \D\gamma' \\
&=&   \int_{\M} \int_{\M} \int_{\P(\gamma,\gamma')}
\overline{g\psi}(\gamma) \phi(\gamma')\, e^{iS(f)} \D f\D\gamma\D\gamma' \\
&=&    \langle g\psi,\phi\rangle_{phys}  \ean
for any $\psi,\phi$.   Here we are assuming
\[          e^{iS(\tilde gf)}\, \D(\widetilde gf) =  e^{iS(f)} \D f,\]
which is one of the expected invariance properties of the path
integral.
It follows that $\I$ includes the space $\J$, the closure of the
span of all vectors of
the form $\psi - g\psi$.  We can therefore define the (spatially)
``diffeomorphism-invariant state space'' $\H_{diff}$ by
$\H_{diff} = \H_{kin} /\J $
and obtain $\H_{phys}$ as a Hilbert space completion
of $\H_{diff}/\K$, where $\K$ is the image of $\I$ in $\H_{diff}$.

To summarize, we obtain the physical state space from the kinematical
state space by taking two quotients:
\ban         \H_{kin} &\to&  \H_{kin}/\J \;\; =\, \H_{diff} \\
             \H_{diff} &\to&  \H_{diff}/\K \hookrightarrow \H_{phys}. \ean
As usual in canonical quantum gravity and topological quantum field
theory, there is no Hamiltonian; instead, all the information about
dynamics is contained in the physical inner product.   The reason, of
course, is that the path integral, which in traditional quantum field
theory described time evolution, now describes the physical inner product.
The quotient map $\H_{diff} \to \H_{phys}$, or equivalently its kernel
$\K$, plays the role of a ``Hamiltonian constraint.''     The quotient map
$\H_{kin} \to \H_{diff}$, or equivalently its kernel $\J$, plays the role
of the ``diffeomorphism constraint,'' which is independent of the dynamics.
(Strictly speaking, we should call $\K$ the ``dynamical constraint,'' as we
shall see that it expresses constraints on the initial data other than
those usually called the Hamiltonian constraint, such as the ``Mandelstam
constraints'' arising in gauge theory.)

It is common in canonical quantum gravity to proceed in a slightly
different manner than we have done here, using subspaces at certain points
where we use quotient spaces \cite{RS,RS2}. For example, $\H_{diff}$ may be
defined as the subspace of $\H_{kin}$ consisting of states invariant under
the action of $\Diff_0(X)$, and $\H_{phys}$ then defined as the kernel of
certain operators, the Hamiltonian constraints.   The method of working
solely with quotient spaces, has, however, been studied by Ashtekar
\cite{Ash}.

The choice between these different approaches will in the end be
dictated by the desire for convenience and/or rigor.  As a heuristic
guiding principle, however, it is worth noting that
the subspace and quotient space approaches are essentially equivalent if
we assume that the subspace $\I$ is closed in the norm topology on
$\H_{kin}$.   Relative to the kinematical inner product, we can identify
$\H_{diff}$ with the orthogonal complement $\J^\perp$, and similarly identify
$\H_{phys}$ with $\I^\perp$.  From this point of
view we have
\[             \H_{phys} \subseteq \H_{diff} \subseteq \H_{kin}  .\]
Moreover, $\psi \in
\H_{diff}$ if and only if $\psi$ is invariant under the action of
$\Diff_0(X)$ on $\H_{kin}$.
To see this, first note that if $g\psi = \psi$ for all $g\in
\Diff_0(X)$, then for all $\phi \in \H_{kin}$ we have
\[         \langle \psi, g\phi - \phi \rangle =
\langle g^{-1}\psi - \psi, \phi \rangle = 0 \]
so $\psi \in \J^{\perp}$.   Conversely, if $\psi \in \J^\perp$,
\[        \langle \psi, g\psi - \psi\rangle = 0 \]
so $\langle \psi,\psi \rangle = \langle\psi,g\psi \rangle$, and since
$g$ acts unitarily on $\H_{kin}$ the Cauchy-Schwarz inequality implies
$g\psi = \psi$.

The approach using subspaces is the one with the clearest connection to
knot theory.
An element $\psi \in \H_{kin}$ is function on the space of multiloops.
If $\psi$ is invariant under the action of $\Diff_0(X)$, we call $\psi$
a ``multiloop invariant.''  In particular, $\psi$ defines an ambient
isotopy invariant of links in $X$ when we restrict it to links (which are
nothing but multiloops that happen to be embeddings).   We see therefore that
in this situation the physical states define link invariants.  As a
suggestive example, take $X = S^3$, and take as the Hamiltonian
constraint an operator $H$ on $\H_{diff}$
that has the property described in Figure \ref{3d}.
\begin{figure}
\centering
\setlength{\unitlength}{0.0125in}%
\begin{picture}(280,40)(60,680)
\thicklines
\put( 18,698){\makebox(0,0)[lb]{\raisebox{0pt}[0pt][0pt]{$H\psi($}}}
\put( 50,720){\vector( 1,-1){ 40}}
\put( 50,680){\vector( 1, 1){ 40}}
\put( 100,698){\makebox(0,0)[lb]
{\raisebox{0pt}[0pt][0pt]{$)\quad =\quad a\psi($}}}

\put(175,720){\vector( 1,-1){ 40}}
\put(200,705){\vector( 1, 1){ 15}}
\put(175,680){\line( 1, 1){ 15}}
\put( 220,698){\makebox(0,0)[lb]{\raisebox{0pt}[0pt][0pt]{$)\; +\; b\psi( $}}}
\put(270,720){\line( 1,-1){ 15}}
\put(295,695){\vector( 1,-1){ 15}}
\put(270,680){\vector( 1, 1){ 40}}

\put( 315,698){\makebox(0,0)[lb]{\raisebox{0pt}[0pt][0pt]{$)\; +\; c\psi( $}}}
\put(368,720){\vector( 1, 0){ 40}}
\put(368,680){\vector( 1, 0){ 40}}
\put( 415,698){\makebox(0,0)[lb]{\raisebox{0pt}[0pt][0pt]{$)$}}}
\end{picture}
\caption[x]{Two-string interaction in 3-dimensional space}
\label{3d}
\end{figure}
Here $a,b,c \in \C$ are arbitrary.   This Hamiltonian constraint represents
the simplest sort of diffeomorphism-invariant two-string interaction in
3-dimensional space.  Defining the physical space $\H_{phys}$ to be the
kernel of $H$, it follows that any $\psi \in H_{phys}$ gives a link
invariant that is just a multiple of the HOMFLY invariant \cite{HOMFLY}.
For appropriate values of the parameters $a,b,c,$ we expect this sort of
Hamiltonian constraint to occur in a generally covariant gauge theory on
4-dimensional spacetime known as $BF$ theory, with gauge group
$SU(N)$ \cite{Horowitz}.  A similar construction working with unoriented
framed multiloops gives rise to the Kauffman polynomial, which is
associated with $BF$ theory with  gauge group $SO(N)$
\cite{Kauffman}.   We see here in its barest form the path from
string-theoretic considerations to link invariants and then to gauge
theory.

In what follows, we start from the other end, and
consider a generally covariant gauge theory on $M$.
Thus we fix a Lie group $G$ and a principal $G$-bundle $P \to M$.
Fixing an identification $M \cong \R \times X$, the
classical configuration space is the space $\A$ of
connections on $P|_{\{0\} \times X}$.  (The physical Hilbert space of
the quantum theory, it should be emphasized, is supposed to be
independent of this identification $M \cong \R \times X$.)
Given a loop $\gamma \maps S^1 \to X$ and a connection $A\in \A$,
let $T(\gamma,A)$ be the corresponding Wilson loop, that is,
the trace of the holonomy of $A$ around $\gamma$ in a fixed
finite-dimensional representation of $G$:
\[           T(\gamma,A) = \tr\, {\rm P}\, e^{ \oint_\gamma A } . \]

The group $\G$ of gauge transformations acts on $\A$. Fix a
$\G$-invariant measure $\D A$
on $\A$ and let $\Fun(\A/\G)$ denote a space of gauge-invariant
functions on $\A$
containing the algebra of functions generated by Wilson loops.   We may
alternatively think of $\Fun(\A/\G)$ as a space of
functions on $\A/\G$ and $\D A$ as a measure on $\A/\G$.
Assume that $\D A$ is invariant under the action of
$\Diff_0(M)$ on $\A/\G$, and define the kinematical state space $\H_{kin}$
to be be the Hilbert space completion of $\Fun(\A/\G)$ in the norm
associated to the kinematical inner product
\[ \langle \psi,\phi\rangle_{kin} = \int_{\A/\G}
\overline\psi(A) \phi(A) \, \D A  .\]

The relation of this kinematical state space and that described above
for a string field theory is given by the loop transform.
Given any multiloop $(\gamma_1, \dots, \gamma_n) \in \M_n$, define the loop
transform $\hat\psi$ of $\psi\in \Fun(\A/\G)$ by
\[            \hat\psi(\gamma_1, \dots, \gamma_n) =
\int_{\A/\G} \psi(A) T(\gamma_1,A) \cdots T(\gamma_n,A) \, \D A  .\]
Take $\Fun(\M)$ to be the space of functions in the range of the loop
transform.  Let us assume, purely for simplicity of exposition, that the
loop transform is one-to-one.  Then we may identify $\H_{kin}$ with
$\Fun(\M)$ just as in the string field theory case.

The process of passing from the kinematical state space to the
diffeomorphism-invariant state space and then the physical state space
has already been treated for a number of generally covariant gauge
theories, most notably quantum gravity \cite{A,Rovelli,RS}.
In order to emphasize the resemblance to the
string field case, we will use a path integral approach.

Fix a time $T > 0$.  Given $A, A' \in \A$,
let $\P(A,A')$ denote the space of connections on
$P|_{[0,T] \times X}$ which restrict to $A$ on $\{0\} \times X$ and
to $A'$ on $\{T\} \times X$.
We assume the existence of a measure on
$\P(A,A')$ which we write as $ e^{iS(a)}
\D a$, using $a$ to denote a connection on $P|_{[0,T] \times X}$.
Again, this generalized measure has some invariance properties
corresponding to the general covariance of the gauge theory.
Define the ``physical'' inner product on $\H_{kin}$ by
\[ \langle \psi,\phi\rangle_{phys} = \int_{\A} \int_{\A} \int_{\P(A,A')}
\overline\psi(A) \phi(A') \, e^{iS(a)} \D a\D A \D A'\]
again assuming that this integral is well-defined and that
$\langle \psi,\psi \rangle \ge 0$ for all $\psi$.
This inner product should be independent of the choice of time $T > 0$.
Letting $\I \subseteq \H_{kin}$ denote the space of norm-zero states,
the physical state space $\H_{phys}$ of the gauge theory is
$\H_{kin}/\I$.  As before, we can use the general covariance of the theory
to show that $\I$ contains the closed span $\J$ of all vectors of
the form $\psi - g\psi$.  Letting $\H_{diff} = \H_{kin}/\J$,
and letting $\K$ be the image of $\I$ in $\H_{diff}$, we again see
that the physical state space is obtained by applying first the
diffeomorphism constraint
\[     \H_{kin} \to \H_{kin}/\J = \H_{diff}   \]
and then the Hamiltonian constraint
\[  \H_{diff} \to  \H_{diff}/\K \hookrightarrow \H_{phys}. \]

In summary, we see that the Hilbert spaces for generally covariant string
theories and generally covariant gauge theories have a similar form,  with
the loop transform relating the gauge theory picture to the string theory
picture.   The key point, again, is that a state $\psi$ in $\H_{kin}$ can
either be regarded as a wavefunction on the classical configuration space
$\A$ for gauge fields, with $\psi(A)$ being the amplitude of a specified
connection $A$, {\it or} as a wavefunction on the classical configuration
space $\M$ for strings, with $\hat\psi(\gamma_1, \cdots, \gamma_n)$ being
the amplitude of a specified $n$-tuple of strings $\gamma_1, \dots,
\gamma_n \maps S^1 \to X$ to be present.   The loop transform depends on
the nonlinear ``duality'' between connections and loops,
\ba              \A/\G \times \M &\to& \C  \nonumber\cr
                 (A, (\gamma_1, \dots, \gamma_n)) &\mapsto&
T(A,\gamma)\cdots T(A,\gamma_n)  \nonumber\ea
which is why we speak of string field/gauge field duality rather than
an isomorphism between string fields and gauge fields.

At this point it is natural to ask what is the difference, apart from
words, between the loop representation of a generally covariant gauge
theory and the sort of purely topological string field theory we have been
considering.   From the Hamiltonian viewpoint (that is, in terms of the
spaces $\H_{kin}$, $\H_{diff}$, and $\H_{phys}$) the difference is not so
great.  The Lagrangian for a gauge theory, on the other hand, is quite a
different object than that of a string field theory. Note that nothing we
have done allows the direct construction of a string field Lagrangian from
a gauge field Lagrangian or vice versa. In the following sections we will
consider some examples:  Yang-Mills theory in 2 dimensions, quantum gravity
in 3 dimensions, and quantum gravity in 4 dimensions.  In no case is a
string field action $S(f)$ known that corresponds to the gauge theory in
question!    However, in 2d Yang-Mills theory a working substitute for the
string field path integral is known: a discrete sum over certain
equivalence classes of maps $f \maps \Sigma \to M$.  This is, in fact, a
promising alternative to dealing with measures on the space $\P$ of string
histories.   In 4 dimensional quantum gravity, such an approach might
involve a sum over ``2-tangles,'' that is, ambient isotopy classes of
embeddings $f \maps \Sigma \to [0,T] \times X$.

\section{Yang-Mills Theory in 2 Dimensions}

We begin with an example in which most of the details have been worked out.
Yang-Mills theory is not a generally covariant theory since it relies for
its formulation on a fixed Riemannian or Lorentzian metric on the spacetime
manifold $M$.  We fix a connected compact Lie group $G$ and a principal
$G$-bundle $P \to M$.  Classically the gauge fields in question are
connections $A$ on $P$, and the Yang-Mills action is given by
\[         S(A) =   -\hf \int_M \tr(F\wedge\star F)   \]
where $F$ is the curvature of $A$ and $\tr$ is the trace in a fixed
faithful unitary representation of $G$ and hence its Lie algebra $\g$.
Extremizing this action we obtain the classical equations of motion, the
Yang-Mills equation
\[           d_A \star F = 0, \]
where $d_A$ is the exterior covariant derivative.

The action $S(A)$ is gauge-invariant so it can be regarded as a function on
the space of connections on $M$ modulo gauge transformations.  The group
$\Diff(M)$ acts on this space, but the action is not
diffeomorphism-invariant.  However, if $M$ is 2-dimensional one may write
$F = f \tensor \om$ where $\om$ is the volume form on $M$ and $f$ is a
section of $P \times_{\Ad} \g$, and then
\[          S(A) = -\hf \int_M \tr(f^2) \,\om.  \]
It follows that the action $S(A)$ is invariant under the subgroup of
diffeomorphisms preserving the volume form $\om$.  So upon quantization one
expects to - and does - obtain something analogous to a topological quantum
field theory, but in which diffeomorphism-invariance is replaced by
invariance under this subgroup. Strictly speaking, then, many of the
results of the previous section  not apply.  In particular, this theory one
has an honest Hamiltonian, rather than a Hamiltonian constraint.  Still, it
illustrates some interesting aspects of gauge field/string field duality.

The Riemannian case of 2d Yang-Mills theory has been extensively
investigated.    An equation for the vacuum expectation values of Wilson
loops for the theory on Euclidean $\R^2$  was found by Migdal
\cite{Migdal}, and these expectation values were explicitly calculated by
Kazakov \cite{Kazakov}.  These calculations were made rigorous using
stochastic differential equation techniques by L.\ Gross, King and Sengupta
\cite{GKS}, as well as Driver \cite{Driver}.  The classical Yang-Mills
equations on Riemann surfaces were extensively investigated by Atiyah and Bott
\cite{AB}, and the quantum theory on Riemann surfaces has been studied by
Rusakov \cite{Rusakov}, Fine \cite{Fine} and Witten \cite{Witten}.    In
particular, Witten has shown that the quantization of 2d Yang-Mills theory
gives a mathematical structure very close to that of a topological quantum
field theory, with a Hilbert space $Z(S^1 \cup \cdots \cup S^1)$ associated
to each compact 1-manifold $S^1 \cup \cdots \cup S^1$, and a vector
$Z(M,\alpha) \in Z(\partial M)$ for each compact oriented 2-manifold $M$
with boundary having total area $\alpha = \int_M \om$.

Let us briefly review some of this work while adapting it to Yang-Mills theory
on $\R \times S^1$ with the Lorentzian metric
\[                 g = dt^2 - dx^2,  \]
where $t \in \R$, $x \in S^1$.  This will simultaneously serve as a brief
introduction to the idea of quantizing gauge theories after symplectic
reduction, which will also be important in 3d quantum gravity.  This
approach is an alternative to the path-integral approach of the previous
section, and in some cases is easier to make rigorous.

Any $G$-bundle $P \to \R \times S^1$ is trivial, so we fix a
trivialization and identify a connection on $P$
with a $\g$-valued 1-form on $\R \times S^1$.
The classical configuration space of the theory is the space $\A$ of
connections on $P|_{\{0\} \times S^1}$.  This may be identified with the
space of $\g$-valued 1-forms on $S^1$.   The classical phase space of the
theory is the cotangent bundle $T^\ast \A$.  Note that a tangent vector
$v \in T_A\A$ may be identified with a $\g$-valued 1-forms on
$S^1$.  We may also think of a $\g$-valued 1-form $E$ on $S^1$
as a cotangent vector, using the nondegenerate inner product:
\[         \langle E,v\rangle = - \int_{S^1} \tr(E \wedge \star v) ,\]
We thus regard the
phase space $T^\ast \A$ as the space of pairs $(A,E)$ of $\g$-valued
1-forms on $S^1$.

Given a connection on $P$ solving the Yang-Mills equation we obtain a
point $(A,E)$ of the phase space $T^\ast \A$ as follows: let $A$ be the
pullback of the connection to $\{0\} \times S^1$, and let $E$ be its
covariant time derivative pulled back to $\{0\} \times S^1$.
The pair $(A,E)$ is called the initial data for the solution, and
in physics $A$ is called the vector potential and $E$ the electric field.
The Yang-Mills equation implies a constraint on $(A,E)$, the Gauss law
\[         d_A \star E = 0,\]
and any pair $(A,E)$ satisfying this constraint is the initial data for
some solution of the Yang-Mills equation.    However, this solution is not
unique, due to the gauge-invariance of the equation.  Moreover, the loop
group $\G = C^\infty(S^1,G)$ acts as
gauge transformations on $\A$, and this action lifts naturally to an action on
$T^\ast \A$, given by:
\[       g \maps (A,E) \to (gAg^{-1} + gd(g^{-1}), gEg^{-1}) . \]
Two points in the phase space $T^\ast \A$
are to be regarded as physically equivalent if they differ by
a gauge transformation.

In this sort of situation it is natural to try to construct a smaller, more
physically relevant ``reduced phase space'' using the process of symplectic
reduction.  The phase space $T^\ast \A$ is a symplectic manifold, but the
constraint subspace
\[       \{ (A,E) \vert\;  d_A \star E = 0\} \subset
T^\ast \A   \]
is not.  However, the constraint $d_A \star E$, integrated
against any $f \in C^\infty(S^1,\g)$ as follows,
\[        \int_{S^1} \tr(fd_A \star E) ,\]
gives a function on phase space that generates a Hamiltonian flow
coinciding with a one-parameter group of gauge
transformations.  In fact, all one-parameter subgroups of $\G$ are
generated by the constraint in this fashion.  Consequently,
the quotient of the constraint subspace by $\G$ is again a
symplectic manifold, the reduced phase space.

In the case at hand there is a very concrete description of the reduced
phase space.  First, by basic results on moduli spaces of flat
connections, the ``reduced configuration space''
$\A/\G$ may be naturally identified with $\Hom(\pi_1(S^1),G)/\Ad G$, which
is just $G/\Ad G$.   Alternatively, one can see this quite concretely.
We may first take the quotient of $\A$
by only those gauge transformations that equal the identity at a given point
of $S^1$:
\[      \G_0 = \{ g \in C^\infty(S^1,G) | \; g(0) = 1 \}.\]
This ``almost reduced'' configuration space
$\A/\G_0$ is diffeomorphic to $G$ itself, with an
explicit diffeomorphism taking each equivalence class
$[A]$ to its holonomy around the circle:
\[         [A] \mapsto {\rm P}\, e^{ \oint_{S^1} A } \]
The remaining gauge transformations form the group $\G/\G_0 \cong G$,
which acts on the almost reduced configuration space $G$ by conjugation, so
$\A/\G \cong G/\Ad G$.

Next, writing $E = edx$, the Gauss law says that $e \in C^\infty(S^1,\g)$
is a flat section, hence determined by its value at the basepoint of $S^1$.
It follows that any point $(A,E)$ in the constraint subspace is determined
by $A \in \A$ together with $e(0) \in \g$.  The quotient of the constraint
subspace by $\G_0$, the ``almost reduced'' phase space, is thus identified
with $T^\ast G$.  It follows that the quotient of the constraint subspace
by all of $\G$, the reduced phase space, is identified with $T^\ast
G/\Ad G$.

The advantage of the almost reduced configuration space and phase space is
that they are manifolds.  Observables of the classical theory can be
identified either with functions on the reduced phase space, or functions
on the almost reduced phase space $T^\ast G$ that are constant on the
orbits of the lift of the adjoint action of $G$.
For example, the Yang-Mills Hamiltonian is initially a function on $T^\ast \A$:
\[       H(A,E) = \hf  \langle E,E\rangle  \]
but by the process of symplectic reduction one obtains a corresponding
Hamiltonian on the reduced phase space.   One can, however,
carry out only part of the process of symplectic
reduction, and obtain a Hamiltonian function on the almost reduced phase
space.     This is just the Hamiltonian for a free particle
on $G$, i.e., for any $p \in T^\ast_g G$ it is given by
\[        H(g,p) = \hf \|p\|^2  \]
with the obvious inner product on $T^\ast_g G$.

Now let us consider quantizing 2-dimensional Yang-Mills theory.
What should be the Hilbert space for the quantized theory on
$\R \times S^1$?   As described in the previous section, it is natural to
take $L^2$ of the reduced configuration space $\A/\G$.  (Since the theory
is not generally covariant, the diffeomorphism and Hamiltonian constraints
do not enter; the ``kinematical'' Hilbert space is the physical Hilbert
space.)  However, to define $L^2(\A/\G)$ requires
choosing a measure on $\A/\G = G/\Ad G$.   We will choose the
pushforward of normalized Haar measure on $G$ by the quotient map $G \to
G/\Ad G$.  This measure has the advantage of mathematical elegance.
While one could also argue for it on physical grounds, we prefer to
simply show {\it ex post facto} that it gives an interesting quantum theory
consistent with other approaches to 2d Yang-Mills theory.

To begin with, note that this measure gives a Hilbert space isomorphism
\[        L^2(\A/\G) \cong L^2(G)_{inv} \]
where the right side denotes the subspace of $L^2(G)$ consisting of
functions constant on each conjugacy class of $G$.   Let $\chi_\rho$ denote
the character of an equivalence class $\rho$ of irreducible representations
of $G$.   Then by the Schur orthogonality relations, the set
$\{\chi_\rho\}$ forms an orthonormal basis of $L^2(G)_{inv}$. In fact, the
Hamiltonian of the quantum theory is diagonalized by this basis.   Since
the Yang-Mills Hamiltonian Hamiltonian on the almost reduced phase space
$T^\ast G$ is that of a classical free particle on $G$,  we take the
quantum Hamiltonian to be that for a quantum free particle on $G$:
\[        H = \Delta/2  \]
where $\Delta$ is the (nonnegative) Laplacian on $G$.
When we decompose the regular representation of $G$ into irreducibles,  the
function $\chi_\rho$ lies in the sum of copies of the representation
$\rho$, so
\ba              H\chi_\rho = \hf c_2(\rho) \chi_\rho ,\label{Ham0} \ea
where $c_2(\rho)$ is the quadratic Casimir of $G$ in the
representation $\rho$. Note that the vacuum (the eigenvector of $H$ with
lowest eigenvalue) is the function $1$, which is $\chi_\rho$ for $\rho$ the
trivial representation.

In a sense this diagonalization of the Hamiltonian
completes the solution of Yang-Mills theory on $\R
\times S^1$.   However, extracting the physics from this solution requires
computing expectation values of physically interesting observables.
To take a step in this direction, and to make the connection to string
theory, let us consider the Wilson loop observables.   Recall that given a
based loop $\gamma \maps S^1 \to S^1$, the classical Wilson loop
$T(\gamma,A)$ is defined by
\[      T(\gamma,A) = \tr {\rm P} e^{ \oint_\gamma A } . \]
We may think of $T(\gamma) = T(\gamma, \cdot)$ as a
function on the reduced configuration space $\A/\G$,
but it lifts to a function
on the almost reduced configuration space $G$, and we prefer to think of
it as such.  In the case at hand these Wilson loop observables depend
only on the homotopy class of the loop, because all connections on $S^1$
are flat.  In the string field picture of Section 2, we obtain a
theory in which all physical states have
\[      \psi(\eta_1, \cdots, \eta_n) =  \psi(\gamma_1, \cdots,
\gamma_n)\]
when $\eta_i$ is homotopic to $\gamma_i$ for all $i$.
We will see this again in 3d quantum gravity.  Letting $\gamma_n \maps
S^1 \to S^1$ be an arbitrary loop of winding number $n$, we have
\[     T(\gamma_n, g) = \tr(g^n)  .\]

Since the classical Wilson loop observables are functions on
configuration space, we may quantize them by interpreting them
as multiplication operators acting on $L^2(G)_{inv}$:
\[       (T(\gamma_n)\psi)(g) = \tr(g^n)\psi(g)  .\]
We can also form elements of $L^2(G)_{inv}$ by applying products of
these operators to the vacuum.  Let
\[     | n_1, \dots, n_k \rangle =  T(\gamma_{n_1}) \cdots T(\gamma_{n_k}) 1 \]
The states
$  | n_1, \dots, n_k \rangle$ may also be regarded as states of a string
theory in
which $k$ strings are present, with winding numbers $n_1, \dots, n_k$,
respectively.    For convenience, we define
$|\emptyset\rangle$ to be the vacuum
state.

The resemblance of the ``string states'' $|n_1, \dots, n_k\rangle$
to states in a bosonic Fock space should be clear.
In particular, the $T(\gamma_n)$ are analogous to ``creation
operators.''  However, we do {\it not} generally have a representation
of the canonical commutation relations.  In fact, the string states
do not necessarily span $L^2(G)_{inv}$, although they do in some
interesting cases.  They are never linearly independent, because
the Wilson loops satisfy relations.  One always has
$T(\gamma_0) = \tr(1)$, for example,
and for any particular group $G$ the Wilson loops will satisfy
identities called Mandelstam identities.
For example, for $G = SU(2)$ and taking traces in the fundamental
representation, the Mandelstam identity is
\[        T(\gamma_n)T(\gamma_m) = T(\gamma_{n+m}) + T(\gamma_{n-m}).\]
Note that this implies that
\[      |n,m\rangle = |n+m\rangle + |n-m\rangle, \]
so the total number of strings present in a given state is ambiguous.
In other words, there is no analog of the Fock space ``number
operator'' on $L^2(G)_{inv}$.

String states appear prominently in the work of
Gross, Taylor, Minahan and Polychronakos \cite{GT,MP} on
$SU(N)$ Yang-Mills theory in 2 dimensions as a string theory.  These
authors, however, work primarily with the large $N$ limit of $SU(N)$
Yang-Mills theory, for since the
work of t'Hooft \cite{t'Hooft} it has been clear that $SU(N)$ Yang-Mills
theory simplifies as $N \to \infty$.
In what follows we will use many ideas from these authors,
but give a string-theoretic formula
for the $SU(N)$ Yang-Mills Hamiltonian that is exact for arbitrary $N$,
instead of working in the large $N$ limit.

For the rest of this section we set $G = SU(N)$ and take traces in the
fundamental representation.  In this case the string states do span
$L^2(G)_{inv}$, and all the linear dependencies between string states
are consequences of the following Mandelstam identities \cite{GamTri}.
Given loops $\eta_1, \dots, \eta_k$ in $S^1$, let
\[      M_k(\eta_1, \dots, \eta_k) =
{1\over k!} \sum_{\sigma \in S_k} {\rm sgn}(\sigma) T(g_{j_{11}} \cdots
g_{j_{1n_1}}) \cdots T(g_{j_{k1}} \cdots g_{j_{kn_k}}) \]
where $\sigma$ has the cycle structure $(j_{11} \cdots j_{1n_1}) \cdots
(j_{k1} \cdots j_{kn_k})$.  Then
\[       M_N(\eta, \dots, \eta) = 1  \]
for all loops $\eta$, and
\[       M_{N+1}(\eta_1, \dots, \eta_{N+1}) = 0  \]
for all loops $\eta_i$.
There are also explicit formulas expressing the string states in terms
of the basis $\{\chi_\rho\}$ of characters.
These formulas are based on the classical theory of Young diagrams,
which we shall briefly review.
The importance of this theory
for 2d Yang-Mills theory is clear from the work of Gross
and Taylor \cite{Gross,GT}.  As we shall see, Young diagrams describe
a ``duality'' between the representation theory of $SU(N)$ and of the symmetric
groups $S_n$ which can be viewed as a mathematical reflection of string
field/gauge field duality.

First, note using the Mandelstam identities
that the string states $|n_1, \cdots, n_k\rangle$ with all
the $n_i$ positive (but $k$ possibly equal to zero) span
$L^2(SU(N))_{inv}$.  Thus we will restrict our attention for
now to states of this kind, which we call ``right-handed.''
There is a 1-1 correspondence between
right-handed string states and conjugacy classes of
permutations in symmetric groups, in which
the string state $|n_1, \cdots, n_k\rangle$
corresponds to the conjugacy class $\sigma$ of all permutations with cycles
of length $n_1, \cdots, n_k$.
Note that $\sigma$ consists of permutations in $S_{n(\sigma)}$, where
$n(\sigma) = n_1 + \cdots + n_k$.   To take advantage of this
correspondence, we simply define
\[       |\sigma\rangle = |n_1, \cdots, n_k\rangle  .\]
when $\sigma$ is the conjugacy class of permutations with cycle lengths
$n_1, \dots, n_k$.   We will assume without loss of generality that
$n_1 \ge \cdots \ge n_k > 0$.

The rationale for this
description of string states as
conjugacy classes of permutations is in fact quite simple.  Suppose we
have length-minimizing strings in $S^1$
with winding numbers $n_1, \dots,n_k$.  Labelling each strand of string
each time it crosses the point $x = 0$, for a total of $n = n_1 + \cdots
+ n_k$ labels, and following the strands around counterclockwise to $x =
2\pi$, we obtain a permutation of the labels, hence an element of $S_n$.
However, since the labelling was arbitrary, the string state really only
defines a conjugacy class $\sigma$ of elements of $S_n$.

In a Young diagram one draws a conjugacy class $\sigma$ with cycles of length
$n_1 \ge \cdots \ge n_k > 0$ as a
diagram with $k$ rows of boxes, having
$n_i$ boxes in the $i$th row.   (See Figure 2.)
\begin{figure}
\centering
\begin{picture}(300,200)(-170,-100)
\thicklines
\put(-120,80){\line(1,0){175}}
\put(-120,80){\line(0,-1){150}}
\put(-120,55){\line(1,0){175}}
\put(55,80){\line(0,-1){25}}
\put(30,80){\line(0,-1){25}}
\put(-120,30){\line(1,0){125}}
\put(5,80){\line(0,-1){50}}
\put(-20,80){\line(0,-1){50}}
\put(-120,5){\line(1,0){75}}
\put(-120,-20){\line(1,0){75}}
\put(-45,80){\line(0,-1){100}}
\put(-120,-45){\line(1,0){50}}
\put(-120,-70){\line(1,0){50}}
\put(-70,80){\line(0,-1){150}}
\put(-95,80){\line(0,-1){150}}
\end{picture}
\caption[x]{Young Diagram}
\label{fig1}
\end{figure}
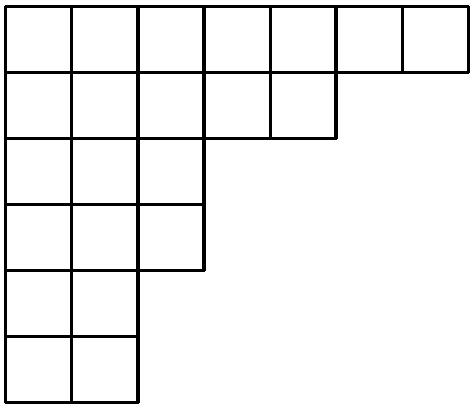
Let $Y$ denote the set of Young diagrams.

On the one hand, there is a map from Young diagrams to
equivalence classes of irreducible representations of $SU(N)$.
Given $\rho \in Y$, we form an irreducible
representation of $SU(N)$, which we also call $\rho$,
by taking a tensor product of $n$ copies of the fundamental
representation, one copy
for each box, and then antisymmetrizing over all copies in each
column and symmetrizing over all copies in each row.
This gives a 1-1 correspondence between Young diagrams with $<
N$ rows and irreducible representations of $SU(N)$.
If $\rho$ has $N$ rows it is equivalent to a representation
coming from a Young diagram having $< N$ rows, and if
$\rho$ has $> N$ rows it is zero-dimensional.
We will write $\chi_\rho$ for the character of the representation
$\rho$; if $\rho$ has $> N$ rows $\chi_\rho = 0$.

On the other hand, Young diagrams with $n$ boxes are in 1-1
correspondence with
irreducible representations of $S_n$.
This allows us to write the Frobenius
relations expressing the string states $|\sigma\rangle$ in terms
of characters $\chi_\rho$ and vice versa.
Given $\rho
\in Y$, we write $\tilde \rho$ for the corresponding representation of
$S_n$.  We define the function $\chi_{\tilde \rho}$
on $S_n$ to be zero for $n(\rho) \ne n$, where $n(\rho)$ is the number
of boxes in $\rho$.  Then the Frobenius relations are
\ba     | \sigma \rangle = \sum_{\rho \in Y}
 \chi_{\tilde\rho}(\sigma)  \chi_\rho ,\label{Frob1} \ea
and conversely
\ba   \chi_\rho = {1\over n(\rho)!}
\sum_{\sigma\in S_{n(\rho)}} \chi_{\tilde\rho}(\sigma)\, |\sigma\rangle .
\label{Frob2} \ea

The Yang-Mills Hamiltonian has a fairly simple description in terms of
the basis of characters $\{\chi_\rho\}$.
First, recall that equation (\ref{Ham0}) expresses the
Hamiltonian in terms of the Casimir.  There is an explicit formula
for the value of the $SU(N)$ Casimir in the representation $\rho$:
\[          c_2(\rho) = Nn(\rho) - N^{-1} n(\rho)^2 +
{n(\rho)(n(\rho)-1)\chi_{\tilde\rho}(``2")\over \dim(\tilde \rho)}  \]
where $``2"$ denotes the conjugacy class of permutations in
$S_{n(\rho)}$ with one cycle of length 2 and the rest of length 1.
It follows that
\ba      H = \hf (N H_0 - N^{-1} H_0^2 + H_1)  \label{Ham1}  \ea
where
\ba    H_0\, \chi_\rho =  n(\rho)\chi_\rho  \label{Ham2}  \ea
and
\ba   H_1\, \chi_\rho = {n(\rho)(n(\rho)-1)\chi_\rho(``2")\over
\dim(\tilde\rho)}
\chi_\rho.  \label{Ham3}  \ea

To express the operators $H_0$ and $H_1$ in string-theoretic terms,
it is convenient to define
string annihilation and creation operators satisfying the canonical
commutation relations.   As noted above, there is no natural way to do
this in $L^2(SU(N))_{inv}$ since the string states
are not linearly independent.  The work of Gross,
Polychronakos relies on the fact that any finite set of distinct
string states becomes linearly independent, in fact orthogonal, for
sufficiently large $N$.  We will proceed slightly differently,
simply {\it defining} a space in which all
the string states are independent.  Let $\H$ be a Hilbert space
having an orthonormal basis
$\{\Chi_\rho\}_{\rho \in Y}$ indexed by all Young diagrams.
For each $\sigma \in Y$,
{\it define} a vector $|\sigma\rangle$ in $\H$ by the Frobenius
relation (\ref{Frob1}).   Then a calculation using the Schur orthogonality
equations twice shows that these string states
$|\sigma\rangle$ are, not only linearly independent, but orthogonal:
\ba    \langle \sigma| \sigma'\rangle &=&
\sum_{\rho,\rho' \in Y} \overline \chi_{\tilde\rho}(\sigma)
\chi_{\tilde\rho'}(\sigma') \langle \Chi_\rho, \Chi_{\rho'} \rangle
\nonumber\cr
&=&  \sum_{\rho\in Y} \overline \chi_{\tilde\rho}(\sigma)
\chi_{\tilde\rho}(\sigma')  \nonumber\cr
&=& {n(\sigma)!\over |\sigma|} \delta_{\sigma \sigma'}  .\nonumber\ea
where $|\sigma|$ is the number of elements in
$\sigma$ regarded as a conjugacy class in $S_n$.
One can also {\it derive} the Frobenius relation (\ref{Frob2})
from these definitions and express the basis $\{\Chi_\rho\}$ in terms of the
string states:
\[   \Chi_\rho = {1\over n(\rho)!}
\sum_{\sigma\in S_{n(\rho)}} \chi_{\tilde\rho}(\sigma) |\sigma\rangle .
\]
It follows that the string states form a basis for $\H$.

The Yang-Mills Hilbert space $L^2(SU(N))_{inv}$ is a quotient space of
the string field Hilbert space $\H$, with the quotient map
\[             j \maps \H \to L^2(SU(N))_{inv}  \]
being given by
\[             \Chi_\rho \mapsto \chi_\rho  .\]
This quotient map sends the string state $|\sigma\rangle$ in $\H$ to the
corresponding string state $|\sigma\rangle \in L^2(SU(N))_{inv}$.
It follows that this quotient map is precisely that which identifies any
two string states that are related by the Mandelstam identities.
It was noted some time ago by Gliozzi and Virasoro \cite{GV}
that Mandelstam identities on string states are strong
evidence for a gauge field interpretation of a string field theory.
Here in fact we will show that the Hamiltonian on the Yang-Mills
Hilbert space $L^2(SU(N))_{inv}$ lifts to a Hamiltonian on $\H$
with a simple interpretation in terms of string interactions, so that
2-dimensional $SU(N)$ Yang-Mills theory is isomorphic to a quotient of
a string theory by the Mandelstam identities.  In the framework of the
previous section, the Mandelstam identities would appear as part of the
``dynamical constraint'' $\K$ of the string theory.

Following equations (\ref{Ham1}-\ref{Ham3}),
we define a Hamiltonian $H$ on the string field Hilbert space $\H$ by
\[      H = \hf(N H_0 - N^{-1} H_0^2 + H_1)  \]
where
\[    H_0 \Chi_\rho =  n(\rho)\Chi_\rho  ,\qquad
  H_1 \Chi_\rho =  {n(n-1)\chi_{\tilde\rho}(``2")\over \dim(\tilde\rho)}
\Chi_\rho.  \]
This clearly has the property that
\[             H j = j H ,\]
so the Yang-Mills dynamics is the quotient of the string field
dynamics.
On $\H$ we can introduce creation operators $a_j^\ast$ ($j > 0$) by
\[       a_j^\ast |n_1, \dots, n_k\rangle =  |j, n_1,
\cdots, n_k\rangle  ,\]
and define the
annihilation operator $a_j$ to be the adjoint of $a_j^\ast$.
These satisfy the following commutation relations:
\[     [a_j,a_k] = [a^\ast_j, a^\ast_k] = 0 ,\qquad
     [a_j, a^\ast_k] = j\delta_{jk}  .\]
We could eliminate the factor of $j$ and obtain the usual canonical
commutation relations by a simple rescaling, but it is more convenient
not to.   We then claim that
\[         H_0 = \sum_{j > 0}  a^\ast_j a_j  \]
and
\[         H_1 = \sum_{j,k > 0}  a^\ast_{j+k} a_j a_k +
a^\ast_j a^\ast_k a_{j+k}  .\]
These follow from calculations by Minahan and Polychronakos
\cite{MP}, which we briefly sketch here.  The Frobenius relations
and the definition of $H_0$ give
\ba          H_0 |\sigma\rangle = n(\sigma) |\sigma\rangle ,
\label{Ham4} \ea
and this implies the formula for $H_0$ as a sum of harmonic oscillator
Hamiltonians $a^\ast_j a_j$.  Similarly, the Frobenius relations and
the definition of $H_1$ give
\[      H_1 |\sigma\rangle =  \sum_{\rho\in Y} {n(\sigma)(n(\sigma)-1)\over
\dim(\tilde \rho)} \chi_{\tilde\rho}(``2") \chi_{\tilde\rho}(\sigma)
\chi_\rho  .\]
Since there are $n(n-1)/2$ permutations $\tau \in S_{n(\sigma)}$ lying in
the conjugacy class $``2"$, we may rewrite this as
\[      H_1 |\sigma\rangle =  \sum_{\rho\in Y, \tau \in ``2"}
{2\over \dim(\tilde \rho)}
\chi_{\tilde\rho}(\sigma) \chi_{\tilde\rho}(\tau) \chi_\rho  .\]
Since
\[        \sum_{\tau \in ``2"} {1\over \dim(\tilde\rho)}
\chi_{\tilde\rho}(\sigma) \chi_{\tilde\rho}(\tau) =
\sum_{\tau \in ``2"} \chi_{\tilde\rho}(\sigma\tau)  \]
the Frobenius relations give
\ba      H_1 |\sigma\rangle =  2\sum_{\tau \in ``2"}
|\sigma\tau\rangle  .\label{Ham5}  \ea
An analysis of the effect of composing $\sigma$ with all possible
$\tau \in ``2"$ shows that either one cycle of $\sigma$ will be broken
into two cycles, or two will be joined to form one, giving the
expression above for $H_1$ in terms of annihilation and creation
operators.
%factor of 2???

We may interpret the Hamiltonian in terms of strings
as follows.  By equation (\ref{Ham4}),
$H_0$ can be regarded as a ``string
tension'' term, since if we represent a string state $|n_1, \dots,
n_k\rangle$ by length-minimizing loops, it is an eigenvector of $H_0$
with eigenvalue equal to $n_1 + \cdots + n_k$, proportional to
the sum of the lengths of the loops.

\begin{figure}
\centering
\setlength{\unitlength}{0.012500in}%
\begin{picture}(95,20)(35,755)
\thicklines
\put( 65,775){\vector( 1,-1){ 20}}
\put( 65,755){\vector( 1, 1){ 20}}
\put(132,775){\vector( 1, 0){ 20}}
\put(132,755){\vector( 1, 0){ 20}}
\put(157,760){\makebox(0,0)[lb]{\raisebox{0pt}[0pt][0pt]{$) $}}}
\put( 28,760){\makebox(0,0)[lb]{\raisebox{0pt}[0pt][0pt]{$H_1\psi($}}}
\put( 90,760){\makebox(0,0)[lb]{\raisebox{0pt}[0pt][0pt]{$) = \psi($}}}
\end{picture}
\caption[x]{Two-string interaction in 1-dimensional space}
\end{figure}

By equation (\ref{Ham5}),
$H_1$ corresponds to a
two-string interaction as in Figure 3.
In this figure only the $x$ coordinate is to be taken seriously; the other
has been introduced only to keep track of the identities of the
strings.   Also, we have switched to treating states as functions on the
space of multiloops.
As the figure indicates, this kind of interaction
is a 1-dimensional version of that which gave the HOMFLY invariant
of links in 3-dimensional space in the previous section.  Here,
however, we have a true Hamiltonian rather than a Hamiltonian constraint.

Figure 3 can also be regarded as two frames of a ``movie'' of
a string worldsheet in 2-dimensional spacetime.  Similar movies
have been used by Carter and Saito to describe string worldsheets in
4-dimensional spacetime \cite{Carter-Saito}.
If we draw the string worldsheet corresponding to
this movie we obtain a surface with a branch point.  Indeed,
in the path integral approach of Gross and Taylor this
kind of term appears in the partition function as part of a sum over
string histories, associated to those histories with branch points.
They also show that the $H_0^2$ term corresponds to string worldsheets
with handles.  When considering the $1/N$ expansion of the
theory, it is convenient to divide the Hamiltonian $H$ by $N$, so that
it converges to $H_0$ as $N \to \infty$.
Then the $H_0^2$ term is proportional to $1/N^2$.
This is in accord with the observation by t'Hooft
\cite{t'Hooft} that in an expansion of the free energy
(logarithm of the partition function) as a power series in $1/N$,
string worldsheets of genus $g$ give terms proportional to $1/N^{2-2g}$.

{} From the work of Gross and Taylor it is also clear that in addition
to the space $\H$ spanned by
right-handed string states one should also consider a space with a
basis of ``left-handed'' string states $|n_1, \cdots, n_k\rangle$ with
$n_i < 0$.  The total Hilbert space of the string theory
is then the tensor product
$\H_+ \tensor \H_-$ of right-handed and left-handed state spaces.  This
does not describe any new states in the Yang-Mills theory per se, but it
is more natural from the string-theoretic point of view.  It follows
from the work of Minahan and Polychronakos that there is a
Hamiltonian $H$ on $\H_+ \tensor \H_-$ naturally described in terms of
string interactions and a quotient map $j \maps \H_+ \tensor \H_- \to
L^2(SU(N))_{inv}$ such that $Hj = jH$.

\section{Quantum Gravity in 3 dimensions}

Now let us turn to a more sophisticated model, 3-dimensional  quantum
gravity.   In 3 dimensions, Einstein's equations say simply that the
spacetime metric is flat, so there are no local degrees of freedom. The
theory is therefore only interesting on topologically nontrivial
spacetimes.  Interest in the mathematics of this theory increased when
Witten \cite{Witten2} reformulated it as a Chern-Simons theory.  Since
then, many approaches to the subject have been developed, not all
equivalent \cite{Carlip}.   We will follow Ashtekar, Husain, Rovelli,
Samuel and Smolin \cite{A,AHRSS} and treat 3-dimensional quantum gravity
using the ``new variables'' and the loop transform, and indicate some
possible relations to string theory.  It is important to note that  there
are some technical problems with the loop transform in  Lorentzian quantum
gravity, since the gauge group is then noncompact \cite{Marolf}.  These are
presently being addressed by Ashtekar and Loll \cite{ALoll} in the
3-dimensional case,  but for simplicity of presentation we will sidestep
them by working with the Riemannian case, where the gauge group is $SO(3)$.

It is easiest to describe the various action principles for gravity using
the abstract index notation popular in general relativity, but we will
instead translate them into language that may be more familiar to
mathematicians, since this seems not to have been done yet.
In this section we describe the ``Witten action,''
applicable to the 3-dimensional case; in the next section we describe the
``Palatini action,'' which applies to any dimension, and the ``Ashtekar
action,'' which applies to 4 dimensions.  The relationship between these
action principles has been discussed rather thoroughly by Peldan
\cite{Peldan}.

Let the spacetime $M$ be an orientable 3-manifold.  Fix a real vector
bundle $\T$ over $M$ that is isomorphic to - but {\it not} canonically
identified with - the tangent bundle $TM$, and fix a Riemannian metric
$\eta$ and an orientation on $\T$.  These define a ``volume form'' $\eps$ on
$\T$, that is, a nowhere vanishing section of $\Lambda^3 \T^\ast$.
The basic fields of the theory are then taken to be a metric-preserving
connection $A$ on $\T$, or ``$SO(3)$ connection,'' together with a
$\T$-valued 1-form $e$ on $M$.   Using the isomorphism
$\T \cong \T^\ast$ given by the metric, the curvature $F$ of $A$ may be
identified with a $\Lambda^2 \T$-valued 2-form.  It follows that the wedge
product $e \wedge F$ may may be defined as a $\Lambda^3 \T$-valued 3-form.
Pairing this with $\eps$ to obtain an ordinary 3-form and then integrating
over spacetime, we obtain the Witten action
\[       S(A,e) = \hf \int_M \eps(e \wedge F)  .\]

The classical equations of motion obtained by extremizing this action are
\[            F = 0 \]
and
\[            d_A e = 0 .\]
Note that we can pull back the metric $\eta$ on $E$ by $e \maps TM \to \T$
to obtain a ``Riemannian metric'' on $M$, which, however, is only
nondegenerate when $e$ is an isomorphism.  When $e$ is an isomorphism we
can also use it to pull back the connection to a metric-preserving
connection on $TM$.  In this case, the equations of motion say simply that
this connection is the Levi-Civita connection of the metric on $M$,
and that the metric on $M$ is flat.
The formalism involving the fields $A$ and $e$ can thus be regarded as a
device for extending the usual Einstein equations in 3 dimensions to the
case of degenerate ``metrics'' on $M$.

Now suppose that $M = \R \times X$, where $X$ is a compact oriented
2-manifold.  The classical configuration and phase spaces and their
reduction by gauge transformations are reminiscent of those for 2d
Yang-Mills theory.  There are, however, a number of subtleties, and we
only present the final results.   The
classical configuration space can be taken as the space $\A$ of
metric-preserving connections on $\T|X$, which we call $SO(3)$ connections
on $X$.  The classical phase space is then the cotangent bundle $T^\ast \A$.
Note that a tangent vector $v \in T_A\A$ is a
$\Lambda^2 \T$-valued 1-form on $X$.  We can thus regard a $\T$-valued
1-form $\tilde E$ on $X$ as a cotangent vector by means of the pairing
\[         \tilde E(v) = \int_X \eps(\tilde E \wedge v) .\]
Thus given any solution $(A,e)$ of the
classical equations of motion, we can pull back $A$
and $e$ to the surface $\{0\} \times X$ and get an $SO(3)$ connection and a
$\T$-valued 1-form on $X$, that is, a point in the phase space $T^\ast \A$.
This is usually written $(A,\tilde E)$, where $\tilde E$ plays a role
analogous to the electric field in Yang-Mills theory.

The classical equations of motion imply constraints on
$(A,\tilde E) \in T^\ast \A$ which define a reduced phase space.
These are the Gauss law, which in this context is
\[   d_A \tilde E = 0,\]
and the vanishing of the curvature $B$ of the connection $A$ on $\T|X$,
which is analogous to the magnetic field:
\[     B = 0.\]
The latter constraint subsumes both the diffeomorphism and Hamiltonian
constraints of the theory.
The reduced phase space for the theory turns out to be
 $T^\ast(\A_0/\G)$, where $\A_0$ is the
space of flat $SO(3)$ connections on $X$, and $\G$ is the group of
gauge transformations \cite{AHRSS}.
As in 2d Yang-Mills theory, it will be attractive quantize after
imposing constraints, taking the physical state space of the
quantized theory to be $L^2$ of the reduced configuration space, if we can
find a tractable description of $\A_0/\G$.

A quite concrete description of $\A_0/\G$ was given by Goldman
\cite{Goldman}.  The moduli space $\F$ of flat $SO(3)$-bundles
has two connected components,
corresponding to the two isomorphism classes of $SO(3)$ bundles on
$M$.  The component corresponding to the bundle $\T|X$ is precisely the space
$\A_0/\G$, so we wish to describe this component.

There is a natural identification
\[        \F \cong  \Hom(\pi_1(X),SO(3))/\Ad(SO(3)), \]
given by associating to any flat bundle the holonomies around
(homotopy classes of) loops.  Suppose that $X$ has
genus $g$.  Then the group $\pi_1(X)$ has a
presentation with $2g$ generators $x_1,y_1, \dots, x_g,y_g$ satisfying
the relation
\[    R(x_i,y_i) = (x_1y_1x_1^{-1}y_1^{-1}) \cdots
(x_gy_gx_g^{-1}y_g^{-1}) = 1 .\]
An element of $\Hom(\pi_1(X), SO(3))$ may thus be identified with
a collection \break
$u_1,v_1, \dots, u_g,v_g$ of elements of $SO(3)$, satisfying
\[    R(u_i,v_i) = 1 ,\]
and a point in $\F$ is an equivalence class $[u_i, v_i]$ of such collections.

The two isomorphism classes of $SO(3)$ bundles on $M$ are
distinguished by their second Stiefel-Whitney number $w_2 \in \Z_2$.
The bundle $\T|X$ is trivial so $w_2(\T|X) = 0$
We can calculate $w_2$ for any point $[u_i, v_i] \in \F$ by the
following method.  For all the elements $u_i, v_i \in SO(3)$, choose lifts
$\tilde u_i, \tilde v_i$ to the universal cover $\widetilde{SO}(3) \cong
SU(2)$.  Then
\[     (-1)^{w_2} = R(\tilde u_i ,\tilde v_i) .\]
It follows that we may think of points of $\A_0/\G$
as equivalence classes of $2g$-tuples
$(u_i,v_i)$ of elements of $SO(3)$ admitting lifts $\tilde u_i, \tilde
v_i$ with
\[     R(\tilde u_i, \tilde v_i) = 1,\]
where the equivalence relation is given by the adjoint action of $SO(3)$.

In fact $\A_0/\G$ is, not a manifold, but a singular variety.
This has been investigated by Narasimhan and Seshadri \cite{NS}, and
shown to be
dimension $d = 6g - 6$ for $g \ge 2$, or $d = 2$ for $g = 1$ (the case
$g = 0$ is trivial and will be excluded below).
As noted,
it is natural to take $L^2(\A_0/\G)$ to be the physical state space, but
but to define this one must choose a measure on $\A_0/\G$.
As noted by Goldman \cite{Goldman}, there is a symplectic structure
$\Om$ on $\A_0/\G$ coming from the following 2-form on $\A_0$:
\[         \Om(B,C) = \int_X \tr(B \wedge C),  \]
in which we identify the tangent vectors $B,C$ with $\End(\T|X)$-valued
1-forms.   The $d$-fold wedge product $\Om \wedge \cdots \wedge \Om$
defines a measure $\mu$ on $\A_0/\G$, the Liouville measure.
On the grounds of elegance and diffeomorphism-invariance
it is customary to use this
measure to define the physical state space $L^2(\A_0/\G)$.

It would be satisfying if there were a string-theoretic interpretation of
the inner product in $L^2(\A_0/\G)$ along the lines of Section 2.     Note
that we may define ``string states'' in this space as follows.
Given any loop $\gamma$ in $X$, the Wilson loop observable $T(\gamma)$ is a
multiplication operator on $L^2(\A_0/\G)$ that only depends on the homotopy
class of $\gamma$.   As in the case of 2d Yang-Mills theory, we can form
elements of $L^2(\A_0/\G)$ by applying products of these operators to the
function $1$, so given $\gamma_1, \cdots, \gamma_k \in \pi_1(X)$, define
\[     | \gamma_1, \dots, \gamma_k \rangle =  T(\gamma_1) \cdots
T(\gamma_k) 1 \]
The first step towards a string-theoretic interpretation of 3d quantum
gravity would be a formula for inner products of the form
\[    \langle\gamma_1, \dots, \gamma_k| \gamma'_1, \dots,
\gamma'_{k'}\rangle, \]
or, equivalently, for integrals of the form
\[     \int_{\A_0/\G} T(\gamma_1,A) \cdots T(\gamma_k,A) d\mu(A) .\]
The author has been unable to find such a formula in the literature except
for the case $g = 1$.
Note that this sort of integral makes sense taking $\A_0$ to be the space of
flat connections for a trivial $SO(N)$ bundle over $X$, for any $N$.
Alternatively, one could formulate 3d quantum gravity as a theory of $SU(2)$
connections and then generalize to $SU(N)$.
One might expect that, as in 2d Yang-Mills theory, the situation
simplifies in the $N \to \infty$ limit.  Ideally, one would like a formula
for
\[    \langle\gamma_1, \dots, \gamma_k| \gamma'_1, \dots,
\gamma'_{k'}\rangle \]
in the $N \to\infty$ limit, together with a method of treating the
finite $N$ case by imposing Mandelstam identities.   In the $N \to \infty$
limit one would also hope for a formula in
terms of a sum over ambient isotopy classes
of surfaces $f \maps \Sigma \to [0,T] \times X$ having
the loops $\gamma_i, \gamma'_i$ as boundaries.

Before concluding this section,
it is worth noting another generally covariant gauge theory in 3 dimensions,
Chern-Simons theory.  Here one fixes an arbitrary Lie group $G$
and a $G$-bundle $P \to M$ over
spacetime, and the field of the theory
is a connection $A$ on $P$.  The action is given by
\[           S(A) = {k\over 4\pi}
 \int \tr(A \wedge dA + {2\over 3}A \wedge A \wedge A)  .\]
As noted by
Witten \cite{Witten2},  3d quantum gravity as we have described it is
essentially the same Chern-Simons theory with gauge group $ISO(3)$, the
Euclidean group in 3 dimension, with the $SO(3)$ connection and triad field
appearing as two parts of an $ISO(3)$ connection.  There is a profound
connection between Chern-Simons theory and knot theory, first demonstrated
by Witten \cite{Witten}, and then elaborated by many researchers (see, for
example, \cite{Atiyah}).  This theory does not quite fit our formalism
because in it the space $\A_0/\G$ of flat connections modulo gauge
transformations plays the role of a phase space, with the Goldman
symplectic structure, rather than a configuration space.  Nonetheless,
there are a number of clues that Chern-Simons theory admits a reformulation
as a generally covariant string field theory.   In fact, Witten has given
such an interpretation using open strings and the Batalin-Vilkovisky
formalism \cite{Witten3}.  Moreover, for the gauge groups $SU(N)$ Periwal
has expressed the partition function for Chern-Simons theory on $S^3$, in
the $N\to\infty$ limit, in terms of integrals over moduli spaces of Riemann
surfaces.  In the case $N = 2$ there is also, as one would expect, an
expression for the vacuum expectation value of Wilson loops, at least for
the case of a link (where it is just the Kauffman bracket invariant), in
terms of a sum over surfaces having that link as boundary \cite{Carter}.
It would be very worthwhile to reformulate Chern-Simons theory as a string
theory at the level of elegance with which one can do so for 2d Yang-Mills
theory, but this has not yet been done.

\section{Quantum Gravity in 4 dimensions}

We begin by describing the Palatini and Ashtekar actions for general
relativity.    As in the previous section, we will sidestep certain problems
with the loop transform by working with Riemannian rather than Lorentzian
gravity.   We shall then discuss some recent work on making the loop
representation rigorous in this case, and indicate some mathematical issues
that need to be explored to arrive at a string-theoretic interpretation of the
theory.

Let the spacetime $M$ be an orientable $n$-manifold.    Fix a bundle $\T$
over $M$ that is isomorphic to $TM$, and fix a Riemannian metric $\eta$ and
orientation on $\T$.   These define a nowhere vanishing section $\eps$ of
$\Lambda^n \T^\ast$.   The basic fields of the theory are then taken to be
a metric-preserving connection $A$ on $\T$, or ``$SO(n)$ connection,'' and
a $\T$-valued 1-form $e$.  We require, however, that $e \maps TM \to \T$ be
a bundle isomorphism; its inverse is called a ``frame field.''  The metric
$\eta$ defines to an isomorphism $\T \iso \T^\ast$ and allows us to identify
the curvature $F$ of $A$ with a section of the bundle
\[         \Lambda^2 \T \tensor \Lambda^2 T^\ast M .\]
We may also regard $e^{-1}$ as a section of $\T^\ast \tensor TM$ and define
$e^{-1} \wedge e^{-1}$ in the obvious manner as a section of the bundle
\[          \Lambda^2 \T^\ast \tensor \Lambda^2 TM .\]
The natural pairing between these bundles gives rise to a function
$F(e^{-1} \wedge e^{-1})$ on $M$.
Using the isomorphism $e$, we can push forward $\eps$
to a volume form $\om$ on $M$.
The Palatini action for Riemannian gravity is then
\[        S(A,e) = \hf \int_M F(e^{-1} \wedge e^{-1})\, \om .\]

We may use the isomorphism $e$ to transfer the metric $\eta$ and
connection $A$ to a metric and connection on the tangent bundle.  Then the
classical equations of motion derived from the Palatini action say
precisely that this connection is the Levi-Civita connection of the metric,
and that the metric satisfies the vacuum Einstein equations (i.e., is Ricci
flat).

In 3 dimensions, the Palatini action reduces to the Witten action, which
however is expressed in terms of $e$ rather than $e^{-1}$.   In 4
dimensions the Palatini action can be rewritten in a somewhat similar form.
Namely, the wedge product $e \wedge e \wedge F$ is a $\Lambda^4 \T$-valued
4-form, and pairing it with $\eps$ to obtain an ordinary 4-form we have
\[       S(A,e) = \hf\int_M \eps(e \wedge e \wedge F) .\]

The Ashtekar action depends upon the fact that in 4 dimensions the metric
and orientation on $\T$ define a Hodge star operator
\[           \ast \maps \Lambda^2 \T \to \Lambda^2 \T \]
with $\ast^2 = 1$ (not to be confused with the Hodge star operator on
differential forms).   This allows us to write $F$ as a sum  $F_+
+ F_-$ of self-dual and anti-self-dual parts:
\[\ast F_{\pm} = \pm F_{\pm}.\]
The remarkable fact is that the action
\[        S(A,e) = \hf \int_M  \eps(e \wedge e \wedge F_+) \]
gives the same equations of motion as
the Palatini action.  Moreover, suppose $\T$ is trivial, as is
automatically the case when $M \cong \R \times X$.
Then $F$ is just an $\so(4)$-valued 2-form on
$M$, and its decomposition into self-dual and anti-self-dual parts
corresponds to the decomposition $\so(4) \cong \so(3) \oplus \so(3)$.
Similarly, $A$ is an $\so(4)$-valued 1-form, and may  thus be written as a
sum $A_+ + A_-$ of ``self-dual'' and ``anti-self-dual'' connections, which
are 1-forms having values in the two copies of $\so(3)$.   It is easy to
see that $F_+$ is the curvature of $A_+$.  This allows us to  regard
general relativity as the theory of a self-dual connection $A_+$ and a
$\T$-valued 1-form  $e$ - the so-called ``new variables'' - with the Ashtekar
action
\[        S(A_+,e) = \hf \int_M \eps(e \wedge e \wedge F_+) .\]

Now suppose that $M = \R \times X$, where $X$ is a compact oriented
3-manifold.  We can take the
classical configuration space to be space $\A$ of right-handed connections
on $\T|X$, or equivalently (fixing a trivialization of $\T$),
$\so(3)$-valued 1-forms on $X$.  A tangent vector $v \in T_A\A$ is thus an
$\so(3)$-valued 1-form, and an $\so(3)$-valued 2-form $\tilde E$ defines a
cotangent vector by the pairing
\[        \tilde E(v) = \int_X \tr(\tilde E\wedge v) .\]
A point in the classical phase space $T^\ast \A$ is thus a pair $(A,\tilde
E)$ consisting of an $\so(3)$-valued 1-form $A$ and an $\so(3)$-valued
2-form $\tilde E$ on $X$.
In the physics literature it is more common to use the natural isomorphism
\[         \Lambda^2 T^\ast X \cong TX \tensor \Lambda^3 T^\ast X \]
given by the interior product to regard the ``gravitational electric
field'' $\tilde E$ as an $\so(3)$-valued vector density, that is, a section
of $\so(3) \tensor TX \tensor \Lambda^3 T^\ast X .$

A solution $(A_+,e)$ of the classical equations of motion determines a
point $(A,\tilde E) \in T^\ast \A$ as follows.   The ``gravitational vector
potential'' $A$ is simply the
pullback of $A_+$ to the surface $\{0\} \times X$.  Obtaining $\tilde E$
from $e$ is a somewhat subtler affair.  First, split the bundle $\T$ as the
direct sum of a 3-dimensional bundle $^3\T$ and a line bundle.
By restricting to $TX$ and then projecting down to $^3\T|X$, the map
\[      e \maps TM \to \T \]
gives a map
\[       ^3 e \maps TX \to \hbox{ $^3\T|X$}, \]
called a ``cotriad field'' on $X$.  Since there is a natural isomorphism of
the fibers of $^3\T$ with $\so(3)$, we may also regard this as an
$\so(3)$-valued 1-form on $X$.  Applying the Hodge star operator we obtain
the $\so(3)$-valued 2-form $\tilde E$.

The classical equations of motion imply constraints on $(A,\tilde E) \in
T^\ast \A$.  These are the Gauss law
\[          d_A \tilde E = 0, \]
and the diffeomorphism and Hamiltonian constraints.  The latter two are
most easily expressed if we treat $\tilde E$ as an $\so(3)$-valued
vector density.    Letting $B$ denote the ``gravitational magnetic field,'' or
curvature of the connection $A$, the diffeomorphism constraint is given
by
\[      \tr \,i_{\tilde E} B = 0 \]
and the Hamiltonian constraint is given by
\[      \tr\, i_{\tilde E} i_{\tilde E} B = 0 .\]
Here the interior product $i_{\tilde E} B$ is defined using $3\times 3$ matrix
multiplication and is a $M_3(\R) \tensor \Lambda^3 T^\ast X$-valued 1-form;
similarly, $i_{\tilde E} i_{\tilde E} B$ is a $M_3(\R) \tensor \Lambda^3
T^\ast X \tensor \Lambda^3 T^\ast X$-valued function.

In 2d Yang-Mills theory and 3d quantum gravity one can impose enough
constraints before quantizing to obtain a finite-dimensional reduced
configuration space, namely the space $\A_0/\G$ of flat connections modulo
gauge transformations.  In 4d quantum gravity this is no longer the case,
so a more sophisticated strategy, first devised by Rovelli
and Smolin \cite{RS}, is required.  Let us first sketch this without
mentioning the
formidable technical problems.   The Gauss law constraint generates gauge
transformations so one forms the reduced phase space $T^\ast(\A/\G)$.
Quantizing, one obtains the kinematical Hilbert space $\H_{kin} =
L^2(\A/\G)$.  One then applies the loop transform and takes
$\H_{kin} = \Fun(\M)$ to be a space of functions of multiloops in $X$.  The
diffeomorphism constraint generates the action of $\Diff_0(X)$ on $\A/\G$,
so in the quantum theory one takes $\H_{diff}$ to be the subspace of
$\Diff_0(X)$-invariant elements of $\Fun(\M)$.  One may then either attempt
to represent the Hamiltonian constraint as operators on $\H_{kin}$, and
define the image of their common kernel in $\H_{diff}$ to be the physical
state space $\H_{phys}$, or attempt to represent the Hamiltonian constraint
directly as operators on $\H_{diff}$ and define the kernel to be
$\H_{phys}$.  (The latter approach is still under development by Rovelli
and Smolin \cite{RS2}.)

Even at this formal level, the full space $\H_{phys}$ has not yet been
determined.  In their original work, Rovelli and Smolin \cite{RS} obtained
a large set of physical states corresponding to ambient isotopy classes of
links in $X$.  More recently, physical states have been constructed from
familiar link such as the Kauffman bracket and certain coefficients of the
Alexander polynomial to all of $\M$.   Some recent developments along these
lines have been reviewed by Pullin \cite{Pullin}.  This approach makes use
of the connection between 4d quantum gravity with cosmological constant and
Chern-Simons theory in 3 dimensions.   It is this work that suggests a
profound connection between knot theory and quantum gravity.

There are, however, significant problems with turning all of this work into
rigorous mathematics, so at this point we shall return to where we left off
in Section 2 and discuss some of the difficulties. In Section 2 we were
quite naive concerning many details of analysis - deliberately so, to
indicate the basic ideas without becoming immersed in technicalities.   In
particular, one does not really expect to have interesting
diffeomorphism-invariant measures on the space $\A/\G$ of connections
modulo gauge transformations in this case.  At best, one expects the
existence of  ``generalized measures'' sufficient for integrating a limited
class of functions.

In fact, it is possible to go a certain distance without becoming involved
with these considerations.   In particular,
the loop transform can be rigorously defined without fixing a
measure or generalized measure on $\A/\G$ if one uses, not the Hilbert
space formalism of the previous section, but a C*-algebraic formalism.  A
C*-algebra is an algebra $A$ over the complex numbers with a norm and an
adjoint or $\ast$ operation satisfying
\[    (a^\ast)^\ast = a,\quad
(\lambda a)^\ast = \overline \lambda a^\ast, \quad (a + b)^\ast = a^\ast +
b^\ast, \quad (ab)^\ast = b^\ast a^\ast ,\]
\[    \|ab\| \le \|a\| \|b\| , \quad \|a^\ast a\| = \|a\|^2 \]
for all $a,b$ in the algebra and $\lambda
\in \C$.  In the C*-algebraic approach to physics, observables are
represented by self-adjoint elements of $A$, while states are elements
$\mu$ of the dual $A^\ast$ that are positive, $\mu(a^\ast a) \ge 0$,
and normalized, $\mu(1) = 1$.   The number $\mu(a)$ then
represents the expectation value of the observable $a$ in the state $\mu$.
The relation to the more traditional Hilbert space approach to quantum
physics is given by the Gelfand-Naimark-Segal (GNS) construction.  Namely,
a state $\mu$ on $A$ defines an ``inner product'' that may however be
degenerate:
\[             \langle a,b \rangle = \mu(a^\ast b)  .\]
Let $I \subseteq A$ denote the subspace of norm-zero states. Then $A/I$
has an honest inner product and we let $\H$ denote the Hilbert space
completion of $A/I$ in the corresponding norm.  It is then easy to check
that $I$ is a left ideal of $A$, so that $A$ acts by left multiplication on
$A/I$, and that this action extends uniquely to a representation of $A$ as
bounded linear operators on $\H$.  In particular, observables in $A$ give
rise to self-adjoint operators on $\H$.

A C*-algebraic approach to the loop transform and generalized measures on
$\A/\G$ was introduced by Ashtekar and Isham \cite{AI} in the context of
$SU(2)$ gauge theory, and subsequently developed by Ashtekar, Lewandowski,
and the author \cite{AL,Baez}.  The basic concept is that of the holonomy
C*-algebra. Let $X$ be a manifold, ``space,'' and let $P \to X$ be a
principal $G$-bundle over $X$.  Let $\A$ denote the space of smooth
connections on $P$, and $\G$ the group of smooth gauge transformations.
Fix a finite-dimensional representation $\rho$ of $G$ and define Wilson
loop functions $T(\gamma) = T(\gamma,\cdot)$ on $\A/\G$ taking traces in
this representation.

Define the ``holonomy algebra'' to be the algebra of functions on
$\A/\G$ generated by the functions $T(\gamma) = T(\gamma,\cdot)$.
If we assume that $G$ is compact and $\rho$ is unitary, the functions
$T(\gamma)$ are bounded and continuous (in the $C^\infty$ topology on
$\A/\G$).  Moreover, the pointwise complex conjugate $T(\gamma)^\ast$ equals
$T(\gamma^{-1})$, where $\gamma^{-1}$ is the orientation-reversed
loop.  We may thus complete the holonomy algebra in the sup norm topology:
\[          \|f\|_\infty = \sup_{A \in \A/\G} |f(A)|  \]
and obtain a C*-algebra of bounded continuous functions on $\A/\G$,
the ``holonomy C*-algebra,'' which we denote as $\Fun(\A/\G)$ in order
to make clear the relation to the previous section.

While in what follows we will assume that $G$ is compact and $\rho$ is
unitary, it is important to emphasize that  for Lorentzian quantum gravity
$G$ is not compact!  This presents important problems in the loop
representation of both 3- and  4-dimensional quantum gravity.   Some
progress in solving these problems has recently been made by Ashtekar,
Lewandowski, and Loll \cite{AL,ALoll,Loll}.

Recall that in the previous section the loop transform of functions on
$\A/\G$ was defined using a measure on $\A/\G$.
It turns out to be more natural to define the loop transform not on
$\Fun(\A/\G)$ but on its dual, as this involves no arbitrary choices.
Given $\mu \in \Fun(\A/\G)^\ast$ we define its loop transform
$\hat\mu$ to be the function on the space $\M$ of multiloops given by
\[        \hat\mu(\gamma_1, \cdots, \gamma_n) =
 \mu(T(\gamma_1) \cdots T(\gamma_n))  .\]
Let $\Fun(\M)$ denote the range of the loop transform.  In favorable
cases, such as $G = SU(N)$ and $\rho$ the fundamental representation,
the loop transform is one-to-one, so
\[  \Fun(\A/\G)^\ast \cong \Fun(\M). \]
This is the real justification for the term ``string field/gauge field
duality.''

We may take the ``generalized measures'' on $\A/\G$ to be simply
elements $\mu \in \Fun(\A/\G)^\ast$, thinking of the pairing $\mu(f)$ as
the integral of $f \in \Fun(\A/\G)$.  If $\mu$ is a state on $\Fun(\A/\G)$, we
may construct the kinematical Hilbert space $\H_{kin}$
using the GNS construction.  Note that the kinematical
inner product
\[          \langle [f], [g] \rangle_{kin} = \mu(f^\ast g)  \]
then generalizes the $L^2$ inner product used in
the previous section.  Note that a choice of generalized measure $\mu$
also allows us to define the loop transform as a linear map from
$\Fun(\A/\G)$ to $\Fun(\M)$
\[        \hat f(\gamma_1, \cdots, \gamma_n) =
\mu(T(\gamma_1) \cdots T(\gamma_n)f)  \]
in a manner generalizing that of the previous section.  Moreover,
there is a unique inner product on $\Fun(\M)$ such that this map
extends to a map from $\H_{kin}$ to the Hilbert space completion of
$\Fun(\M)$.
Note also that $\Diff_0(X)$ acts on $\Fun(\A/\G)$
and dually on $\Fun(\A/\G)^\ast$.  The kinematical Hilbert space
constructed from a $\Diff_0(X)$-invariant state $\mu \in \Fun(\A/\G)$
thus becomes a unitary representation of $\Diff_0(X)$.

It is thus of considerable interest to find a more concrete description of
$\Diff_0(X)$-invariant states on the holonomy C*-algebra $\Fun(\A/\G)$.  In
fact, it is not immediately obvious that any exist, in general! For
technical reasons, the most progress has been made in the real-analytic
case.  That is, we take $X$ to be real-analytic, $\Diff_0(X)$  to consist
of the {\it real-analytic} diffeomorphisms connected to the identity, and
$\Fun(\A/\G)$ to be the holonomy C*-algebra generated by real-analytic
loops.  Here Ashtekar and Lewandowski have constructed a
$\Diff_0(X)$-invariant state on $\Fun(\A/\G)$ that is closely analogous to
the Haar measure on a compact group \cite{AL}.  They have also given a
general characterization of such diffeomorphism-invariant states.  The
latter was also given by the author \cite{Baez}, using a slightly different
formalism, who also constructed many more examples of
$\Diff_0(X)$-invariant states on $\Fun(\A/\G)$.   There is thus some real
hope that the loop representation of generally covariant gauge theories can
be made rigorous in cases other than the toy models of the previous two
sections.

We conclude with some speculative remarks concerning 4d quantum gravity and
2-tangles.  The correct inner product on the physical Hilbert space of 4d
quantum gravity has long been quite elusive.   A path-integral formula for
the inner product  has been investigated recently by Rovelli
\cite{Rovelli}, but there is as yet no manifestly well-defined
expression along these lines. On the other hand, an inner product for
``relative states'' of quantum gravity in the Kauffman bracket state has
been rigorously constructed by the author \cite{BaezTang}, but there are
still many questions about the physics here. The example of 2d Yang-Mills
theory would suggest an expression for the inner product of string states
\[           \langle \gamma_1, \cdots, \gamma_n | \gamma'_1, \cdots,
\gamma'_n\rangle  \] as a sum over ambient isotopy classes of surfaces $f
\maps \Sigma \to [0,T] \times X$ having  the loops $\gamma_i, \gamma'_i$ as
boundaries.   In the case of embeddings, such surfaces are known as
``2-tangles,'' and have been intensively investigated by Carter and Saito
\cite{Carter-Saito} using the technique of ``movies.''

The relationships between 2-tangles, string theory, and the loop
representation of 4d quantum gravity are tantalizing but still rather
obscure.   For example, just as the Reideister moves relate any two
pictures of the same tangle in 3 dimensions, there are a set of movie moves
relating any two movies of the same 2-tangle in 4 dimensions.   These moves
give a set of equations whose solutions would give 2-tangle invariants.
For example, the analog of the Yang-Baxter equation is the Zamolodchikov
equation, first derived in the context of string theory \cite{Zam}.   These
equations can be understood in terms of category theory, since just as
tangles form a braided tensor category, 2-tangles form a braided tensor
2-category \cite{Fischer}.  It is thus quite significant that Crane
\cite{Crane} has initiated an approach to generally covariant field theory
in 4 dimensions using braided tensor 2-categories. This approach also
clarifies some of the significance of conformal field theory for
4-dimensional physics, since braided tensor 2-categories can be constructed
from certain conformal field theories.   In a related development,
Cotta-Ramusino and Martellini \cite{CM} have endeavored to construct
2-tangle invariants from generally covariant gauge theories, much as tangle
invariants may be constructed using Chern-Simons theory.   Clearly it will
be some time before we are able to appraise the significance of all this
work, and the depth of the relationship between string theory and the loop
representation of quantum gravity.


\begin{thebibliography} {10}

\bibitem{A} A.\ Ashtekar, New variables for classical and quantum gravity,
{\sl Phys.\ Rev.\ Lett.\ }{\bf 57} (1986), 2244-2247.

New Hamiltonian formulation of general
relativity, {\sl Phys.\ Rev.\ }{\bf D36} 1587-1602.

{\sl Lectures on Non-perturbative Canonical Quantum
Gravity,} Singapore, World Scientific, 1991.

\bibitem{Ash}  A.\ Ashtekar, unpublished notes, June 1992.

\bibitem{AI}
A.\ Ashtekar and C.\ J.\ Isham, Representations of the holonomy
algebra of gravity and non-abelian gauge theories, {\sl Class.\ and
Quant.\ Grav.\ }{\bf 9} (1992), 1069-1100.

\bibitem{AL}
A.\  Ashtekar and J.\ Lewandowski, Completeness of Wilson loop
functionals on the moduli space of $SL(2,C)$ and $SU(1,1)$
connections, {\sl Class.\ and Quant.\ Grav.\ }{\bf 10} (1993) 673-694.

Representation theory of analytic holonomy C*-Algebras,
this volume.

\bibitem{ALoll} A.\ Ashtekar and R.\ Loll, New loop representations
for 2+1 gravity, Syracuse U.\ preprint.

\bibitem{AHRSS} A.\ Ashtekar, V.\ Husain, C.\ Rovelli, J.\ Samuel and
L.\ Smolin, 2+1 gravity as a toy model for the 3+1 theory, {\sl Class.\
and Quant.\ Grav.\ }{\bf 6} (1989) L185-L193.

\bibitem{Atiyah} M.\ Atiyah, {\sl The Geometry and Physics of Knots,}
Cambridge U.\ Press, Cambridge, 1990.

\bibitem{AB} M.\ Atiyah and R.\ Bott, The Yang-Mills equations over
Riemann surfaces, {\sl Phil.\ Trans.\ Roy.\ Soc.\ London} {\bf A308}
(1983) 523-615.

\bibitem{BaezTang} J.\ Baez, Quantum gravity and the algebra of
tangles, {\sl Jour.\ Class.\ Quant.\ Grav.\ }{\bf 10} (1993) 673-694.

\bibitem{Baez}  J.\ Baez, Diffeomorphism-invariant generalized measures
on the space of connections modulo gauge transformations,
to appear in the proceedings of the Conference on Quantum
Topology, eds.\ L.\ Crane and D.\ Yetter, hep-th/9305045.

Link invariants, functional integration, and holonomy algebras,
U.\ C.\ Riverside preprint, hep-th/9301063.

\bibitem{Carlip} S.\ Carlip, Six ways to quantize (2+1)-dimensional
gravity, U.\ C.\ Davis preprint, gr-qc/9305020.

\bibitem{Carter}  J.\ S.\ Carter, {\sl How Surfaces Intersect in
Space: an Introduction to Topology,} World Scientific,
Singapore, 1993.

\bibitem{Carter-Saito} J.\ S.\ Carter and M.\ Saito, Reidemeister
moves for surface isotopies and their interpretation as moves to movies,
U.\ of South Alabama preprint.

Knotted surfaces, braid movies, and beyond, this volume.

\bibitem{CM}  P.\ Cotta-Ramusino and M.\ Martellini, this
volume.

\bibitem{Crane}  L.\ Crane, Topological field theory as the key to
quantum gravity, this volume.

\bibitem{Driver} B.\ Driver, ${\rm YM_2}$: continuum expectations,
lattice convergence, and lassos, {\sl Comm.\ Math.\ Phys.\ }{\bf 123}
(1989) 575-616.

\bibitem{Fine} D.\ Fine, Quantum Yang-Mills on a Riemann surface, {\sl
Comm.\ Math.\ Phys.\ }{\bf 140} (1991) 321-338.

\bibitem{Fischer}  J.\ Fischer, 2-categories
and 2-knots, Yale U.\ preprint, Feb.\ 1993.

\bibitem{HOMFLY}  P.\ Freyd, D.\ Yetter, J.\ Hoste,
W.\ Lickorish, K.\ Millett, and A.\ Ocneanu,
A new polynomial invariant for
links, {\sl Bull.\ Amer.\ Math.\ Soc.\ }{\bf 12} (1985) 239-246.

\bibitem{GamTri} R.\ Gambini and A.\ Trias, Gauge dynamics in the
C-representation, {\sl Nucl.\ Phys. }{\bf B278} (1986) 436-448.

\bibitem{GerNev} J.\ Gervais and A.\ Neveu, The quantum dual string wave
functional in Yang-Mills theories, Phys.\ Lett.\ {\bf B80}
(1979), 255-258.

\bibitem{GV} F.\ Gliozzi and M.\ Virasoro, The interaction among dual
strings as a manifestation of the gauge group, {Nucl.\ Phys.\ }{\bf
B164} (1980) 141-151.

\bibitem{Goldman} W.\ Goldman, The symplectic nature of fundamental
groups of surfaces, {\sl Adv.\ Math.\ }{\bf 54} (1984) 200-225.

Invariant functions on Lie groups and Hamiltonian flows of surface
group representations, {\sl Invent.\ Math.\ }{\bf 83} (1986) 263-302.

Topological components of spaces of representations, {\sl Invent.\
Math.\ }{\bf 93} (1988) 557-607.

\bibitem{Gross} D.\ Gross, Two dimensional QCD as a string theory,
U.\ C.\ Berkeley preprint, Dec.\ 1992, hep-th/9212149.

\bibitem{GT} D.\ Gross and W.\ Taylor IV, Two dimensional QCD is a
string theory, U.\ C.\ Berkeley preprint, Jan.\ 1993, hep-th/9301068.

Twists and Wilson loops in
the string theory of two dimensional QCD, U.\ C.\ Berkeley preprint,
Jan.\ 1993, hep-th/9303046.

\bibitem{GKS} L.\ Gross, C.\ King, A.\ Sengupta, Two-dimensional
Yang-Mills theory via stochastic differential equations, {\sl Ann.\
Phys.\ }{\bf 194} (1989) 65-112.

\bibitem{Horowitz} G.\ Horowitz, Exactly soluble diffeomorphism-invariant
theories, {\sl Comm.\ Math.\ Phys.\ }{\bf 125} (1989) 417-437.

\bibitem{Kauffman}  L.\ Kauffman, {\sl Knots and Physics,} World
Scientific, Singapore, 1991.

\bibitem{Kazakov} V.\ Kazakov, Wilson loop average for an arbitrary
contour in two-dimensional $U(N)$ gauge theory, {\sl Nuc.\ Phys.\ }{\bf
B179} (1981) 283-292.

\bibitem{Loll} R.\ Loll, J.\ Mour\~ao, and J.\ Tavares,
Complexification of gauge theories, Syracuse U.\ preprint,
hep-th/930142.

\bibitem{MM} Y.\ Makeenko and A.\ Migdal, Quantum chromodynamics as
dynamics of loops, {\sl Nucl.\ Phys.\ }{\bf B188} (1981) 269-316.

Loop dynamics: asymptotic
freedom and quark confinement, {\sl Sov.\ J.\ Nucl.\ Phys.\ }{\bf 33} (1981)
882-893.

\bibitem{Marolf} D.\ Marolf, Loop representations for 2+1 gravity
 on a torus, Syracuse University preprint, March 1993, gr-qc/9303019.

An illustration of 2+1 gravity loop transform troubles,
Syracuse University preprint, May 1993, gr-qc/9303019.

\bibitem{Migdal}  A.\  Migdal, Recursion equations in gauge field
theories {\sl Sov.\ Phys.\  JETP} {\bf 42} (1975) 413-418.

\bibitem{Minahan} J.\ Minahan, Summing over inequivalent maps in the
string theory interpretation of two dimensional QCD, University of
Virginia preprint, hep-th/9301003

\bibitem{MP} J.\ Minahan and A.\ Polychronakos, Equivalence of two
dimensional QCD and the $c = 1$ matrix model, University of Virginian
preprint, hep-th/9305153.

\bibitem{NRS} S.\ Naculich, H.\ Riggs, and H.\ Schnitzer,
Two-dimensional Yang-Mills theories are string theories, Brandeis U.\
preprint, hep-th/9305097.

\bibitem{Nambu} Y.\ Nambu, QCD and the string model, {\sl Phys.\ Lett.\
}{\bf B80} (1979) 372-376.

\bibitem{NS} M.\ Narasimhan and C.\ Seshadri, Stable and unitary vector
bundles on a compact Riemann surface, {\sl Ann.\ Math.\ }{\bf 82} (1965)
540-567.

\bibitem{Peldan} P.\ Peldan, Actions for gravity, with generalizations:
a review, U.\ of G\"oteborg preprint, May 1993, gr-qc/9305011.

\bibitem{Periwal} V.\ Periwal, Chern-Simons theory as topological closed
string, Institute for Advanced Studies preprint.

\bibitem{Polyakov} A.\ Polyakov, Gauge fields as rings of glue, {\sl
Nucl.\ Phys.\ }{\bf  B164} (1979) 171-188.

{\sl Gauge fields and strings,} Harwood Academic Publishers, Chur, 1987.

\bibitem{Pullin}  J.\ Pullin, Knot theory and quantum gravity in loop
space: a primer, to appear in {\sl Proc. of the Vth Mexican School of
Particles and Fields,} ed.\ J. L. Lucio, World Scientific, Singapore;
preprint available as hep-th/9301028.

\bibitem{Rovelli} C.\ Rovelli, The basis of the
Ponzano-Regge-Turaev-Viro-Ooguri model is the loop representation basis,
Pittsburgh U.\ preprint, April 1993, hep-th/9304164.

\bibitem{RS}  C.\ Rovelli and L.\ Smolin, Loop representation for
quantum general relativity {\sl Nucl.\ Phys.\ }{\bf B331} (1990),
80-152.

\bibitem{RS2}
C.\ Rovelli and L.\ Smolin,
The physical hamiltonian in non-perturbative quantum gravity, Pennsylania
State U.\ preprint, August 1993, gr-qc/9308002.

\bibitem{Rusakov}  B.\ Rusakov, Loop averages and partition functions in
$U(N)$ gauge theory on two-dimensional manifolds, {\em Mod. Phys. Lett.}
{\bf A5}, (1990) 693-703.

\bibitem{t'Hooft} G.\ t'Hooft, A two-dimensional model for mesons, {\sl
Nucl.\ Phys.\ }{\bf B75} (1974), 461-470.

\bibitem{Witten}  E.\  Witten, On quantum gauge theories in two
dimensions {\sl Comm.\ Math.\ Phys.\ }{\bf 141} (1991) 153-209.

Localization in gauge theories, lectures at M.\ I.\ T., February 1992.

\bibitem{Witten2} E.\ Witten, 2+1 dimensional gravity as an exactly
soluble system, {\sl Nucl.\ Phys.\ }{\bf B311} (1988) 46-78.

\bibitem{Witten3} E.\ Witten, Chern-Simons gauge theory as a string theory,
to appear in the Floer Memorial Volume, Institute for Advanced
Studies preprint, 1992.

\bibitem{Zam} A.\ Zamolodchikov, Tetrahedron equations and the
relativistic $S$-Matrix of straight-strings in $2+1$-dimensions,
{\sl Comm.\ Math.\ Phys.\ }{\bf 79}, (1981) 489-505.

\bibitem{Zwiebach} B.\ Zwiebach, Closed string field theory: an
introduction, M.\ I.\ T.\ preprint, May 1993, hep-th/9305026.

\end{thebibliography}
\end{document}